\documentclass[twocolumn]{aastex62}
\usepackage{color,soul}
\usepackage{comment}
\usepackage{amsmath}
\usepackage{tablefootnote}

\newcommand{\oiii}{O{\sc iii}}

\graphicspath{{./}{figures/}{figures/kinematics/}}

\submitjournal{ApJ}

\shorttitle{KCWI Measurements of Void Dwarf Galaxies}
\shortauthors{de los Reyes et al.}

\begin{document}

\title{The Stellar Kinematics of Void Dwarf Galaxies Using KCWI}

\correspondingauthor{Mithi A. C. de los Reyes}
\email{mdlreyes@stanford.edu}

\author[0000-0002-4739-046X]{Mithi A. C. de los Reyes}
\affiliation{Department of Physics, Stanford University, 382 Via Pueblo Mall, Stanford, CA 94305, USA}
\affiliation{Kavli Institute for Particle Astrophysics \& Cosmology, P.O. Box 2450, Stanford University, Stanford, CA 94305, USA}

\author[0000-0001-6196-5162]{Evan N. Kirby}
\affiliation{Department of Physics \& Astronomy, University of Notre Dame, 225 Nieuwland Science Hall, Notre Dame, IN 46556, USA}

\author[0000-0002-1945-2299]{Zhuyun Zhuang}
\affiliation{Department of Astronomy, California Institute of Technology, 1200 E. California Blvd., Pasadena, CA 91125, USA}

\author[0000-0002-2052-822X]{Charles C. Steidel}
\affiliation{Department of Astronomy, California Institute of Technology, 1200 E. California Blvd., Pasadena, CA 91125, USA}

\author[0000-0003-4520-5395]{Yuguang Chen}
\affiliation{Department of Physics \& Astronomy, University of California Davis, 1 Shields Ave, Davis, CA 95616, USA}

\author{Coral Wheeler}
\affiliation{Department of Physics \& Astronomy, California State Polytechnic University Pomona, 3801 West Temple Avenue, Pomona, CA 91768, USA}

\begin{abstract}
Dwarf galaxies located in extremely under-dense cosmic voids are excellent test-beds for disentangling the effects of large-scale environment on galaxy formation and evolution.
We present integral field spectroscopy for low-mass galaxies ($M_{\star}=10^{7}-10^{9}~M_{\odot}$) located inside ($N=21$) and outside ($N=9$) cosmic voids using the Keck Cosmic Web Imager (KCWI).
Using measurements of stellar line-of-sight rotational velocity $v_{\mathrm{rot}}$ and velocity dispersion $\sigma_{\star}$, we test the tidal stirring hypothesis, which posits that dwarf spheroidal galaxies are formed through tidal interactions with more massive host galaxies.
We measure low values of $v_{\mathrm{rot}}/\sigma_{\star}\lesssim2$ for our sample of isolated dwarf galaxies, and we find no trend between $v_{\mathrm{rot}}/\sigma_{\star}$ and distance from a massive galaxy $d_{L^{\star}}$ out to $d_{L^{\star}}\sim10$~Mpc.
These suggest that dwarf galaxies can become dispersion-supported ``puffy'' systems even in the absence of environmental effects like tidal interactions.
We also find indications of an upward trend between $v_{\mathrm{rot}}/\sigma_{\star}$ and galaxy stellar mass, perhaps implying that stellar disk formation depends on mass rather than environment.
Although some of our conclusions may be slightly modified by systematic effects, our main result still holds: that isolated low-mass galaxies may form and remain as puffy systems rather than the dynamically cold disks predicted by classical galaxy formation theory. 
\end{abstract}

\section{Introduction} 
\label{sec:intro}

Galaxies are not closed systems.
Not only do they affect their surroundings---through, e.g., outflows of gas---but their evolution is often affected by their immediate environments.
Low-mass ``dwarf'' galaxies\footnote{The definition of ``dwarf'' galaxies can vary significantly; here, we consider ``classical'' dwarf galaxies with stellar masses $<10^{9}$~$M_{\odot}$.} are thought to be particularly susceptible to environmental effects.
Due to their small gravitational potential wells, they tend to be more strongly influenced by outflows of metal-enriched gas (or inflows of pristine gas) and are therefore more metal-poor than massive galaxies \citep[][]{Tremonti2004}.

One way to disentangle the effect of environment in galaxy evolution is to study systems in which potential environmental effects are minimized.
This requires looking beyond our own neighborhood: even the handful of currently-known Local Group dwarf galaxies that are relatively isolated (i.e., not clearly associated with either the Milky Way or M31) are still within $\sim1$~Mpc of the nearest $L^{\star}$ galaxy\footnote{$L^{\star}$ galaxies are approximately Milky Way-mass galaxies; throughout this work we adopt a relatively conservative definition of $L^{\star}$ galaxies as galaxies with stellar masses $M_{\star}>10^{10}~M_{\odot}$.}\citep{McConnachie2021}.
Even the most distant dwarf galaxies in the Local Group could be ``backsplash galaxies'' that may have interacted with an $L^{\star}$ host in the past \citep{Teyssier2012}.
We instead search for dwarf galaxies in the lowest-density environments in the universe: cosmic voids.
The vast majority of the matter in the universe resides in filaments in the so-called ``cosmic web'' \citep{Bond1996}, but galaxies can still be found in the large ($\gtrsim10$~Mpc) and underdense ($\rho_{\mathrm{void}}\sim0.1\rho_{\mathrm{avg}}$) voids between filaments.
Dwarf galaxies in these cosmic voids can be located $\gtrsim5$~Mpc from other galaxies---perhaps as far removed from the external influences of other galaxies as possible in the local universe---making them an ideal population for observing how galaxies evolve in near-total isolation.

Furthermore, $\Lambda$CDM predicts that the evolution of void galaxies will be slowed relative to ``field'' galaxies located in cosmic filaments, due to longer timescales between galaxy-galaxy interactions \citep{Goldberg2004}. 
Indeed, most void dwarfs appear to be optically blue, compact, and actively star forming, suggesting that they are less evolved than their field counterparts \citep[e.g.,][]{Kreckel2012,Grogin2000,Grogin1999}.
This may make void dwarf galaxies useful analogs to the high-redshift galaxies that were the building blocks for Milky Way-like galaxies.
Detailed characterizations of high-$z$ analogs in the local universe might provide complementary information to direct high-$z$ observations taken by, e.g., JWST \citep{Gardner2006}.

A number of previous studies have aimed to characterize the galaxies located in voids \citep[e.g.,][and references therein]{Pustilnik2011,Pustilnik2016,Pustilnik2019,Kreckel2012,Kreckel2014,Penny2015,Beygu2016,Beygu2017,Douglass2017,Kniazev2018,Wegner2019,Florez2021,DominguezGomez2022}.
However, the vast majority of these studies have aimed to measure integrated galaxy properties, including global colors, star formation rates, and metallicities.
This paper describes an observational program designed to measure the \emph{spatially-resolved} properties of a sample of local void dwarf galaxies. 
In particular, this program aims to probe the stellar kinematics of these galaxies, which can help shed light on the dynamical processes that drive galaxy formation.

Classical galaxy formation theory suggests that all galaxies form as thin, rotationally supported disks \citep[][]{White1978,Fall1980,Mo1998}.
However, it is not clear whether this picture of disk formation extends to dwarf galaxies ($M_{\star}\lesssim10^{9}~M_{\odot}$).
Disk-like morphologies are not as readily apparent among dwarf galaxies, particularly compared to the obvious and dramatic spiral disks common among higher-mass galaxies, leading some authors to suggest that dwarf and disk galaxies are structurally distinct systems \citep{Schombert2006}.
Detailed kinematic measurements are most readily available for the dwarf galaxies closest to us: our own Local Group.
The Local Group hosts both gas-rich star-forming dwarf irregular (dIrr) galaxies and gas-poor dwarf spheroidal (dSph) galaxies \citep[e.g.,][]{McConnachie2012}.
The prevalence of these two classes is a well-known function of location: most dIrrs live outside the virial radii of the Milky Way or M31, while satellite galaxies are typically dSphs.
This local ``morphology-density relation'' \citep[first noted by][]{Einasto1974} appears to be a direct consequence of environmental effects, as tides and ram-pressure stripping can remove gas from gas-rich dIrrs and turn them into gas-poor dSphs \citep[e.g.,][]{Grcevich2009,Spekkens2014,Putman2021}.

Environmental effects have also been thought to affect the stellar kinematics of Local Group dwarf galaxies.
In the ``tidal stirring'' model \citep{Mayer2001}, a rotationally supported dwarf galaxy with a stellar disk will experience repeated tidal shocks as it passes through the pericenter of its orbit around a massive host galaxy.
These shocks may produce a tidally induced bar, which transfers angular momentum to the outer regions of the galaxy. 
As high-angular momentum material is stripped, the overall rotation of the galaxy decreases, transforming it into a pressure-supported, kinematically ``puffy'' stellar system.
Simulations have had some success in showing that dSphs may be formed from dIrrs through this mechanism \citep[e.g.,][and references therein]{Kazantzidis2017}.

Yet the kinematic distinction between dIrrs and dSphs is perhaps more ambiguous than the tidal stirring model would suggest.
Stellar spectroscopy has revealed that a number of Local Group dIrrs are primarily dispersion-supported \citep[e.g.,][]{Leaman2012,Kirby2017}---and indeed, nearly all the lowest-mass ($M_{*}\lesssim10^{8}~\mathrm{M}_{\odot}$) Local Group dwarf galaxies are either dispersion-supported or only weakly rotationally supported \citep{Kirby2014,Wheeler2017}.
\citet{Wheeler2017} suggested that this may point to a formation scenario in which dwarf galaxies initially form as ``puffy'' stellar systems rather than dynamically cold disky systems, and they show that zoom-in simulations of isolated dwarf irregular galaxies are consistent with this picture.
In this paper, we demonstrate that void dwarf galaxies provide a useful test of these scenarios.

The structure of this paper is as follows.
We describe the observations and data reduction in Section~\ref{sec:data}. 
In Section~\ref{sec:analysis}, we present measurements of spatially-resolved stellar kinematics, before discussing the implications of these results in Section~\ref{sec:discussion}.
We summarize our conclusions in Section~\ref{sec:conclusion}.

\section{Data} 
\label{sec:data}

Since the stellar populations of galaxies outside the Local Group cannot be resolved, spatially-resolved spectroscopy is needed to estimate stellar kinematic properties.
While long-slit spectroscopy can be used to obtain measurements along single axes, integral field units (IFUs) can obtain spectral information across the full spatial extent of an extended source.
We used the Keck Cosmic Web Imager \citep[KCWI;][]{Morrissey2018} on the Keck II telescope to obtain IFU observations of dwarf galaxies inside and outside voids.
In this section, we describe the sample selection, observations, and data reduction process.

\subsection{Sample selection}
\label{sec:sample}

We selected a sample of void dwarf galaxies from Table 1 of \citet{Douglass2018}, who identified $993$ void dwarf galaxies from the Sloan Digital Sky Survey Data Release 7 \citep[SDSS DR7;][]{Abazajian2009}.
To classify these galaxies as ``void'' galaxies, \citet{Douglass2018} used the void catalog compiled by \citet{Pan2012}, which was built from the SDSS DR7 catalog using the VoidFinder algorithm \citep{Hoyle2002,ElAd1997}.
This algorithm finds geometric voids using the spatial distribution of massive galaxies in SDSS DR7 (with absolute magnitudes $M_{r}<-20$).
Isolated galaxies (described as having the third nearest-neighbor more than $\sim7$~Mpc/$h$ away) are removed, then all remaining ``wall'' galaxies are placed on a three-dimensional grid. 
Every grid cell devoid of ``wall'' galaxies is potentially part of a void, so VoidFinder aims to identify the maximal sphere that can be drawn in the void: a sphere is grown from each empty grid cell, reaching its maximum size once four galaxies are present on its surface (since a sphere is uniquely defined by four non-coplanar points).
Overlapping spheres are combined, and any sphere with radius $>10~\mathrm{Mpc}$ is then associated with a void.

We chose $19$ void dwarf galaxies, aiming to evenly span a range of stellar masses from $10^{7}-10^{9}~M_{\odot}$ (i.e., approximately one galaxy in each $0.1~\mathrm{dex}$ stellar mass bin).
Because gas-phase metallicities are a property of particular interest in void dwarf galaxies \citep[e.g.,][]{Kreckel2014,Douglass2017,Pustilnik2016}, we further prioritized galaxies in each mass bin with the highest [\oiii]$\lambda 4363$ fluxes.\footnote{[\oiii]$\lambda 4363$ is the weakest of the emission lines required to measure ``direct'' gas-phase metallicities.}
These preliminary parameters used in our sample selection were obtained from the SDSS value-added catalogs: stellar masses were estimated using the star-forming Portsmouth \citep{Maraston2005} method, which fit stellar evolution models to SDSS photometry, while the emission line fluxes were obtained from the MPA/JHU value-added catalog \citep{Tremonti2004}.
We note that these parameter estimates are only used for target selection, and we compute independent stellar masses from mid-infrared photometry later in our analysis (Section~\ref{sec:mass}).

By targeting galaxies selected from geometric voids in a galaxy survey, this sample selection implicitly assumes that the relatively bright ($m_{r}<17.77$) galaxies in SDSS DR7 perfectly trace the underlying dark matter distribution of the universe.
This may not be the case \citep[see][for a review of galaxy bias]{Desjacques2018}, which means our void sample could be contaminated with galaxies that are not in truly low-density voids---that is, galaxies that live in regions devoid of relatively bright galaxies, but which still contain ``dark'' non-emitting gas, ultra low surface brightness systems, or dark matter.
While a full investigation of this selection is beyond the scope of this paper, we are able to confirm that these void galaxies are at least extremely isolated from massive host galaxies (Section~\ref{sec:dLstar}).

An additional $7$ dwarf galaxies were also selected from SDSS Data Release 16 \citep[SDSS DR16;][]{SDSS16} as a control sample, with stellar masses similar to those of galaxies in the void sample.
These were selected to have relatively low redshift ($z<0.02$), visible [\oiii{}] and H$\beta$ lines in their SDSS spectra, and (as with the void galaxies) were prioritized based on their [\oiii]~$\lambda 4363$ flux.
These control galaxies were either observed or identified as galaxies in SDSS data releases after DR7, so they could not be directly checked against the SDSS DR7 void catalog of \citet{Pan2012}.
Given the relative rarity of void dwarf galaxies---\citet{Douglass2018} identified 993/9519 ($10\%$) of the SDSS DR7 dwarf galaxies as ``void'' galaxies---the majority of the control galaxies are likely non-void galaxies.
However, without further verification, we simply consider these to be ``field'' galaxies (as none of them appear to have nearby massive host galaxies) throughout the remainder of this analysis.

Finally, three dwarf galaxies were added to the sample as potential objects of interest.
These include Pisces A and B, which are nearby dwarf galaxies that have been identified at the boundary between voids and higher-density filaments.
Their star formation histories have undergone a recent increase, potentially triggered by gas accretion from a denser environment, suggesting that they are in the process of exiting the voids in which they likely formed \citep{Tollerud2016}.
IFU maps of Pisces A and B will help identify how this environmental transition affects the kinematic and chemical properties of these galaxies.

The last galaxy in the sample, denoted ``reines65'' in this paper, is ID 65 from \citet{Reines2020}, who recently reported discoveries of luminous compact radio sources in nearby dwarf galaxies that are consistent with radiation from accreting black holes.
Several of these sources appear to be located outside their host galaxies' central regions, and \citet{Reines2020} suggested that these radio sources are evidence for so-called ``wandering'' (i.e., off-nuclear) black holes.
Optical spectroscopy has previously been employed to search for central supermassive black holes in dwarf galaxies \citep[e.g.,][]{Reines2013,Moran2014,Sartori2015}, but the IFU data here present the first opportunity for investigating optical radiation from an \emph{off-nuclear} black hole candidate in a dwarf galaxy.

Table~\ref{tab:voidsample} lists the properties of our final void and control samples.

\begin{deluxetable*}{lllllhl}
\tablecolumns{7} 
\tablecaption{General properties of void and field dwarf galaxy sample. \label{tab:voidsample}} 
\tablehead{ 
\colhead{ID$^{a}$} & \colhead{RA} & \colhead{Dec} & \colhead{$z^{b}$} & \colhead{$g^{b}$} & \nocolhead{$\log M_{\star}^{b}$} & \colhead{Type$^{c}$} \\
\colhead{} & \colhead{(J2000)} & \colhead{(J2000)} & \colhead{} & \colhead{(mag)} & \nocolhead{[$M_{\odot}$]} & \colhead{}
}
\startdata
1180506 & 09 12 51.73 & +31 40 51.48 & 0.0064 & 17.17 & 6.99 & v \\
281238 & 09 45 40.99 & +01 37 03.87 & 0.0064 & 18.38 & 7.06 & v \\
1904061 & 08 48 43.52 & +22 55 47.60 & 0.013 & 17.93 & 7.47 & v \\
821857 & 10 06 42.44 & +51 16 24.23 & 0.0162 & 17.49 & 7.47 & v \\
1158932 & 09 28 44.47 & +35 16 41.14 & 0.0151 & 17.42 & 7.53 & v \\
866934 & 09 16 25.07 & +43 00 19.30 & 0.0085 & 16.66 & 7.57 & v \\
825059 & 08 13 39.49 & +36 42 34.56 & 0.0130 & 16.99 & 8.1 & v \\
2502521 & 09 13 19.89 & +12 32 07.32 & 0.0161 & 18.11 & 7.63 & v \\
1228631 & 10 13 58.42 & +39 48 01.62 & 0.007 & 15.65 & 7.64 & v \\
1876887 & 08 49 56.66 & +25 41 02.61 & 0.008 & 16.67 & 7.65 & v \\
1246626 & 11 30 11.93 & +44 27 16.07 & 0.0172 & 17.06 & 7.68 & v \\
1142116 & 08 18 19.70 & +24 31 36.94 & 0.0073 & 15.56 & 8.06 & v \\
955106 & 10 16 28.21 & +45 19 17.53 & 0.0055 & 14.86 & 8.10 & v \\
1063413 & 11 23 22.03 & +45 45 16.34 & 0.0202 & 17.37 & 8.22 & v \\
1074435 & 09 48 00.79 & +09 58 15.43 & 0.0104 & 16.30 & 8.35 & v \\
1785212 & 07 50 41.62 & +50 57 40.28 & 0.0187 & 17.51 & 8.36 & v \\
1280160 & 11 07 13.71 & +06 24 42.38 & 0.0085 & 15.88& 8.53 & v \\
1782069 & 10 04 38.88 & +67 49 22.05 & 0.0145 & 16.99 & 8.59 & v \\
1126100 & 09 34 03.03 & +11 00 21.67 & 0.0085 & 15.15 & 9.02 & v \\
Pisces A & 00 14 46.00 & +10 48 47.01 & 0.0008 & 17.56 & \ldots & v \\
Pisces B & 01 19 11.70 & +11 07 18.22 & 0.0020 & 17.43 & \ldots & v \\
SDSS J0133+1342 & 01 33 52.56 & +13 42 09.39 & 0.0087 & 17.92& 6.56 & f \\
AGC 112504 & 01 36 40.92 & +15 05 12.14 & 0.0088 & 17.84 & 7.58 & f \\
UM 240 & 00 25 07.43 & +00 18 45.63 & 0.0109 & 17.04 & 7.64 & f \\
SHOC 150 & 03 04 57.97 & +00 57 14.09 & 0.0121 & 17.86 & 7.78 & f \\
LEDA 3524 & 00 58 55.47 & +01 00 17.44 & 0.0179 & 16.47 & 7.99 & f \\
LEDA 101427 & 00 24 25.95 & +14 04 10.65 & 0.0142 & 15.86 & 8.35 & f \\
IC 0225 & 02 26 28.29 & +01 09 37.92 & 0.0051 & 14.06 & 8.72 & f \\	
reines65 & 11 36 42.72 & +26 43 37.68 & 0.0333 & 15.65 & 7.64 & f
\enddata
\tablenotetext{a}{Galaxy ID. For most void galaxies, these are the galaxy index numbers from the KIAS Value Added Galaxy Catalog \citep{Choi2010}. For most field galaxies, these are the galaxy names preferred by NASA/IPAC Extragalactic Database (NED). The galaxy ``reines65'' is identified as galaxy 65 from \citep[][]{Reines2020}.}
\tablenotetext{b}{Redshifts and $g$-band magnitudes are from SDSS.}
\tablenotetext{c}{Galaxy classification as ``v'' for void dwarf galaxy or ``f'' for field dwarf galaxy. Pisces A and B are classified as ``void'' galaxies for our purposes, although they are likely moving into a cosmic filament (see text).}
\end{deluxetable*}

\subsection{Observations and data reduction}
\label{sec:observations}

IFU data were obtained for the sample over 2.5 nights using KCWI, an optical integral field spectrograph on the Nasmyth platform of the 10 m Keck II telescope \citep{Morrissey2018}.
KCWI has multiple configurations; in order to match the typical angular size of the dwarf galaxies in my sample, we used the medium slicer and blue BL grating centered at $\lambda = 4500$~\AA.
This combination yields a $20''\times16.5''$ field of view, nominal spectral resolution of $\sim2.5$~\AA{} ($\sigma\sim71~\mathrm{km}~\mathrm{s}^{-1}$) at $4500$~\AA{}, and a usable wavelength range of $3500-5500$~\AA.

Table~\ref{tab:voidobs} describes the observations of each galaxy. 
For galaxies with multiple exposures, we rotated the position angles by $\pm10^\circ$ for each exposure in order to minimize spatial covariance during stacking.
For each object exposure, we observed a patch of nearby sky with the same exposure time and position angle to perform sky subtraction.
We processed all object exposures using the IDL version of the KCWI data reduction pipeline\footnote{\url{https://github.com/Keck-DataReductionPipelines/KcwiDRP}}, which produces flux-calibrated data cubes.
Data cubes of the same object were then aligned and stacked using a drizzling algorithm as presented in \citet{Chen2021}\footnote{\url{https://github.com/yuguangchen1/kcwi}}.

\begin{deluxetable*}{llllll}
\tablecolumns{5} 
\tablecaption{Observations of void and field dwarf galaxies. \label{tab:voidobs}} 
\tablehead{ 
\colhead{Object} & \colhead{Exposures} & \colhead{Position angles} & \colhead{Date} & \colhead{Airmass} \\
\colhead{} & \colhead{(s)} & \colhead{($^\circ$)} & \colhead{(dd-mm-yy)} & \colhead{}
}
\startdata
\multicolumn{5}{c}{Void dwarf galaxies}\\
\tableline
    1180506 & 2$\times$300 & 167.6, 177.6 & 29-12-2019 & 1.03 \\
	281238 & 2$\times$900 & 139.9, 149.9 & 28-12-2019 & 1.25 \\
	1904061 & 3$\times$800 & 44.1, 54.1, 64.1 & 28-12-2019 & 1.01 \\
	821857 & 2$\times$700 & 97.4, 107.4 & 28-12-2019 & 1.17 \\
	1158932 & 2$\times$700 & 47.7, 57.7 & 29-12-2019 & 1.10 \\
	866934 & 2$\times$300 & 4.4, 14.4 & 29-12-2019 & 1.11 \\
	825059 & 1$\times$600, 2$\times$700, 1$\times$940 & 20.1, 30.1 & 30-12-2019 & 1.20 \\
	2502521 & 2$\times$400 & 18.0, 28.0 & 28-12-2019 & 1.19 \\
		& 2$\times$800 & 8.0, 18.0 & 29-12-2019 & 1.25 \\
	1228631 & 2$\times$600 & 16.0 & 22-01-2020 & 1.06 \\
	1876887 & 2$\times$400 & 145.7, 175.7 & 28-12-2019 & 1.14 \\
	1246626 & 2$\times$800 & 49.0 & 22-01-2020 & 1.10 \\
	1142116 & 2$\times$900 & 144.3, 154.3 & 28-12-2019 & 1.12 \\
	955106 & 1$\times$600 & 9.0 & 22-01-2020 & 1.12 \\
	1063413 & 4$\times$900 & 167.0, 177.0 & 22-01-2020 & 1.26 \\
	1074435 & 1$\times$600, 3$\times$840 & 103.0, 113.0 & 22-01-2020 & 1.28 \\
	1785212 &  2$\times$400 & 50.0, 60.0 & 29-12-2019 & 1.20 \\
		& 2$\times$850 & 50.0, 60.0 & 30-12-2019 & 1.50 \\
	1280160 & 2$\times$700 & 145.0 & 22-01-2020 & 1.09 \\
	1782069 & 4$\times$750 & 37.1, 47.1, 57.1 & 29-12-2019 & 1.62 \\
	1126100 & 2$\times$300 & 149.2, 159.2 & 29-12-2019 & 1.02 \\
	Pisces A & 1$\times$600, 2$\times$1200 & 0.0, 10.0 & 28-12-2019 & 1.11 \\
		& 2$\times$900 & 0.0, 10.0 & 30-12-2019 & 1.35 \\
	Pisces B & 1$\times$600, 2$\times$1000 & 140.0, 150.0 & 30-12-2019 & 1.02 \\
\tableline
\multicolumn{5}{c}{Field dwarf galaxies}\\
\tableline
    SDSS J0133+1342 & 1$\times$500 & 128.0 & 22-01-2020 & 1.11 \\
	AGC 112504 & 2$\times$900 & 134.1, 144.1 & 28-12-2019 & 1.51 \\
	UM 240 & 4$\times$1000 & 91.0, 101.0 & 22-01-2020 & 1.13 \\
	SHOC 150 & 2$\times$600, 2$\times$660 & 118.0, 128.0 & 22-01-2020 & 1.44 \\
	LEDA 3524 & 2$\times$400 & 147.4, 157.4 & 28-12-2019 & 1.55 \\
	LEDA 101427 & 2$\times$400 & 14.8, 24.8 & 30-12-2019 & 1.54 \\
	IC 0225 & 2$\times$900 & 167, 177 & 28-12-2019 & 1.10 \\	
	reines65 & 4$\times$600 & 45.0, 55.0 & 22-01-2020 & 1.10
\enddata
\end{deluxetable*}

\section{Analysis} 
\label{sec:analysis}

The final stacked data cubes are then analyzed using a custom pipeline\footnote{The full pipeline, along with all custom code used in this paper, is available at \url{https://github.com/mdlreyes/void-dwarf-analysis}}.
Although many of the details are tailored for this specific application, the pipeline follows many of the same general steps as other IFS survey analysis pipelines, including the Mapping Nearby Galaxies at Apache Point Observatory Data Analysis Pipeline \citep[MaNGA DAP;][]{Westfall2019} and Pipe3D \citep{Sanchez2016}.

\subsection{Binning and covariance correction}

Some initial steps are taken to prepare the data cubes: the cubes are first corrected for Galactic reddening using the $E(B-V)$ color indices measured by \citet{Schlafly2011}, and the observed wavelength array is divided by $(1+z)$ to correct for redshift.
Each data cube is then spatially binned to increase the continuum signal-to-noise (S/N).
The S/N of an individual spaxel is often too low to reliably fit with stellar continuum templates, particularly in low surface brightness regions like galaxy outskirts or faint dwarf galaxies. 
We therefore increase the S/N by spatially binning spaxels, i.e., averaging multiple adjacent spaxels.
To do this, we use \texttt{vorbin}, an adaptive spatial binning algorithm that produces Voronoi tessellations \citep{Cappellari2003}.

We first define the nominal S/N of an individual spaxel, assuming that each spaxel is independent.
Because we are primarily interested in measuring information from the stellar continuum in each spaxel, we use the formula for the detrended \emph{continuum} S/N defined by \citet{RosalesOrtega2012}:
\begin{equation}
\label{eq:sn-cont}
    \left(\frac{S}{N}\right)_{c} = \frac{\mu_{c}}{\sigma_{c}},
\end{equation}
where $\mu_{c}$ is the mean of the flux in the continuum band $f(\lambda)_{c}$, and $\sigma_{c}$ is the detrended standard deviation (i.e., the standard deviation in the difference between $f(\lambda)_{c}$ and a linear fit to $f(\lambda)_{c}$).
We take $f(\lambda)_{c}$ to be the flux across the continuum range $4750-4800$~\AA, which lacks strong emission features.

The S/N values from Equation~\ref{eq:sn-cont} are likely overestimates, since stacking data cubes introduces covariance between adjacent spaxels.
To account for this, rather than computing full covariance matrices for each spaxel in every data cube, we use an empirical formula to estimate the ratio between the ``true'' noise $\epsilon_{\mathrm{true}}$ and the noise assuming no covariance $\epsilon_{\mathrm{no\,covar}}$.
This ratio, denoted $\eta$, is assumed to be a function of the bin size, with the form suggested by \citet{Husemann2013}:
\begin{equation}
    \eta = \frac{\epsilon_{\mathrm{true}}}{\epsilon_{\mathrm{no\,covar}}} = 
    \begin{cases}
    \beta(1 + \alpha\log N) & N \leq N_{\mathrm{thresh}}\\
    \beta(1 + \alpha\log N_{\mathrm{thresh}}) & N > N_{\mathrm{thresh}}.
    \end{cases}
    \label{eq:epsilonratio}
\end{equation}
Here, $N$ is the number of spaxels in each bin, $\alpha$ describes the strength of the dependence of $\eta$ on bin size, and $\beta$ is a normalization factor. 
Above a threshold bin size $N_{\mathrm{thresh}}$, additional spaxels are assumed to be far enough apart that they do not add any extra covariance, so the ratio $\eta$ is capped at a constant.

\begin{figure}[t!]
    \centering
    \epsscale{1.15}
    \plotone{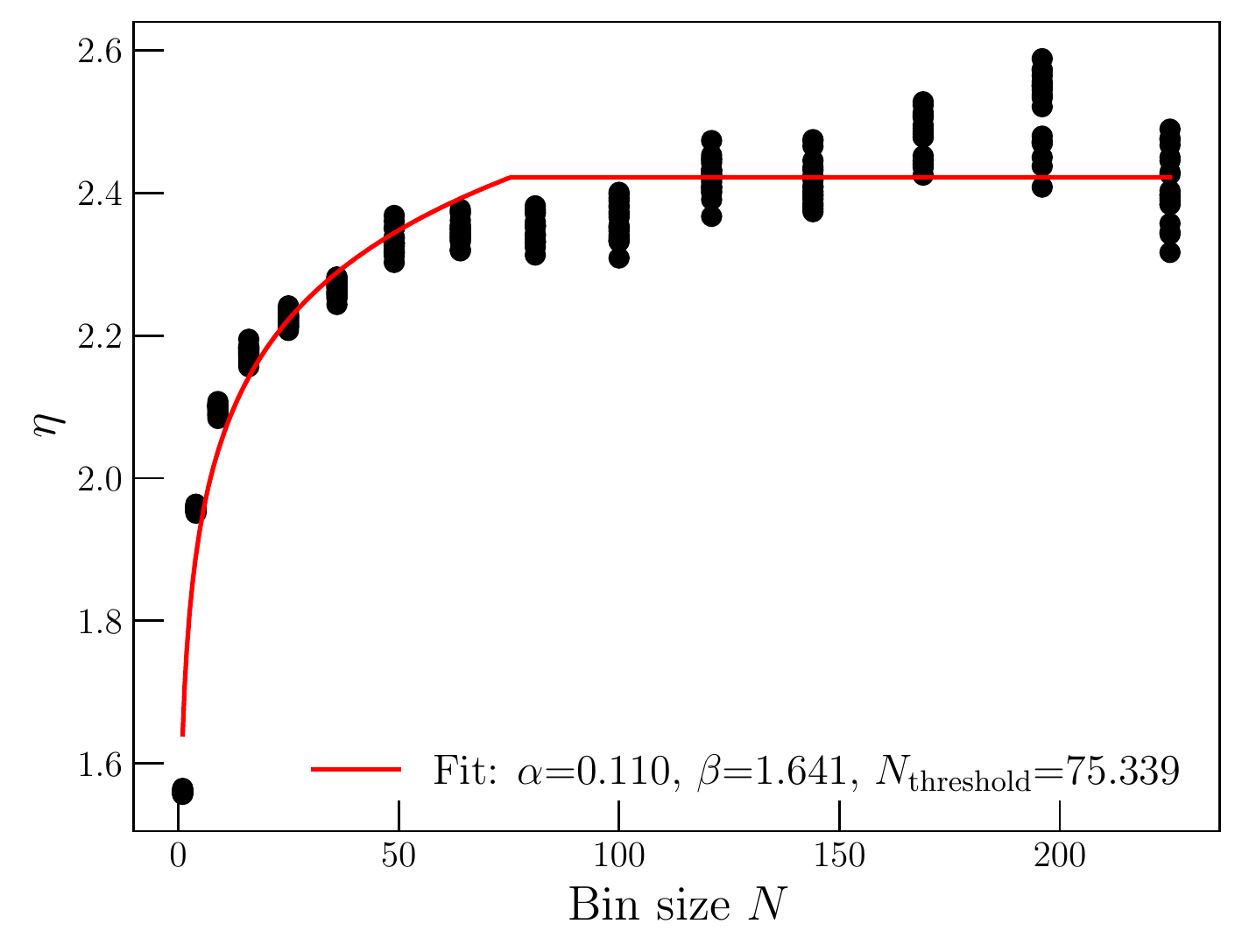}
    \caption{The ratio $\eta = \epsilon_{\mathrm{true}}/\epsilon_{\mathrm{no\,covar}}$ as a function of bin size for an example with 4 stacked exposures. Black points represent empirical estimates from mock data cubes, as described in the text. The red line indicates the best-fit empirical curve of the form Equation~\ref{eq:epsilonratio}.}
    \label{fig:covartest}
\end{figure}

We estimate the free parameters $\{\alpha,\beta,N_{\mathrm{thresh}}\}$ following the procedure of \citet{Law2016} and using the several modules from the CWITools package \citep{OSullivan2020}.
First, we create mock data cubes in which all pixels have fluxes independently drawn from a normal distribution $\mathcal{N}\sim(1,1)$ with mean and variance both unity.
These mock cubes are stacked following the same drizzling procedure as the actual data cubes from individual exposures, producing mock intensity and variance cubes.
For a stacked cube, the spaxels are binned using a simple boxcar of size $n^{2}$ where $n$ varies.
The standard deviation of each bin in the mock intensity cube is an estimate of the ``true'' noise $\epsilon_{\mathrm{true}}$, since it accounts for the effects of stacking.
The stacked mock variance cube, on the other hand, is used to compute a separate noise estimate $\epsilon_{\mathrm{no\,covar}}$ using simple error propagation rules, assuming that each spaxel is independent.
The ratio of these two estimates can be plotted as a function of bin size $N$ and fit with a curve of the functional form described in Equation~\ref{eq:epsilonratio} to determine the best-fit values of $\{\alpha,\beta,N_{\mathrm{thresh}}\}$.
Figure~\ref{fig:covartest} demonstrates an example of this fitting for four stacked exposures, showing that Equation~\ref{eq:epsilonratio} is a good representation of $\eta$.

By multiplying this empirical estimate for $\eta$ by the noise estimate (i.e., dividing Equation~\ref{eq:sn-cont} by $\eta$), the S/N within a bin can be corrected for the effects of spatial covariance.
Using covariance-corrected S/N values, the \texttt{vorbin} algorithm then creates bins with a target S/N while optimally preserving spatial resolution.
The target S/N is at least 10 for all galaxies. 
We attempt to maximize S/N per bin while maintaining spatial resolution; for some galaxies with longer exposures, we set a higher target S/N per bin (for reference, the maximum target S/N is 50 for galaxy 955106).

\subsection{Continuum fitting and kinematics measurements}

Taking the average spectrum in each bin, we now fit the stellar continuum using the full spectral-fitting algorithm \texttt{pPXF} \citep{Cappellari2004,Cappellari2017}.
This algorithm attempts to determine the line-of-sight velocity distribution (LOSVD) of the stars in a galaxy by fitting a galaxy spectrum with a combination of templates. 
We use templates from the MILES stellar library of single stellar population models \citep{SanchezBlazquez2006,FalconBarroso2011,Vazdekis2015}.
These models are produced, assuming a universal \citet{Kroupa2001} stellar initial mass function (IMF), from BaSTI isochrones \citep{Hidalgo2018}, which cover a metallicity range $-2.27<\mathrm{[Fe/H]}<+0.40$ and age range $0.03-14.0$~Gyr. 
The spectra in this library have wavelength range $354-741$~nm with a constant resolution FWHM of 2.51\AA{}.

\begin{figure*}[t!]
     \epsscale{0.8}
     \plotone{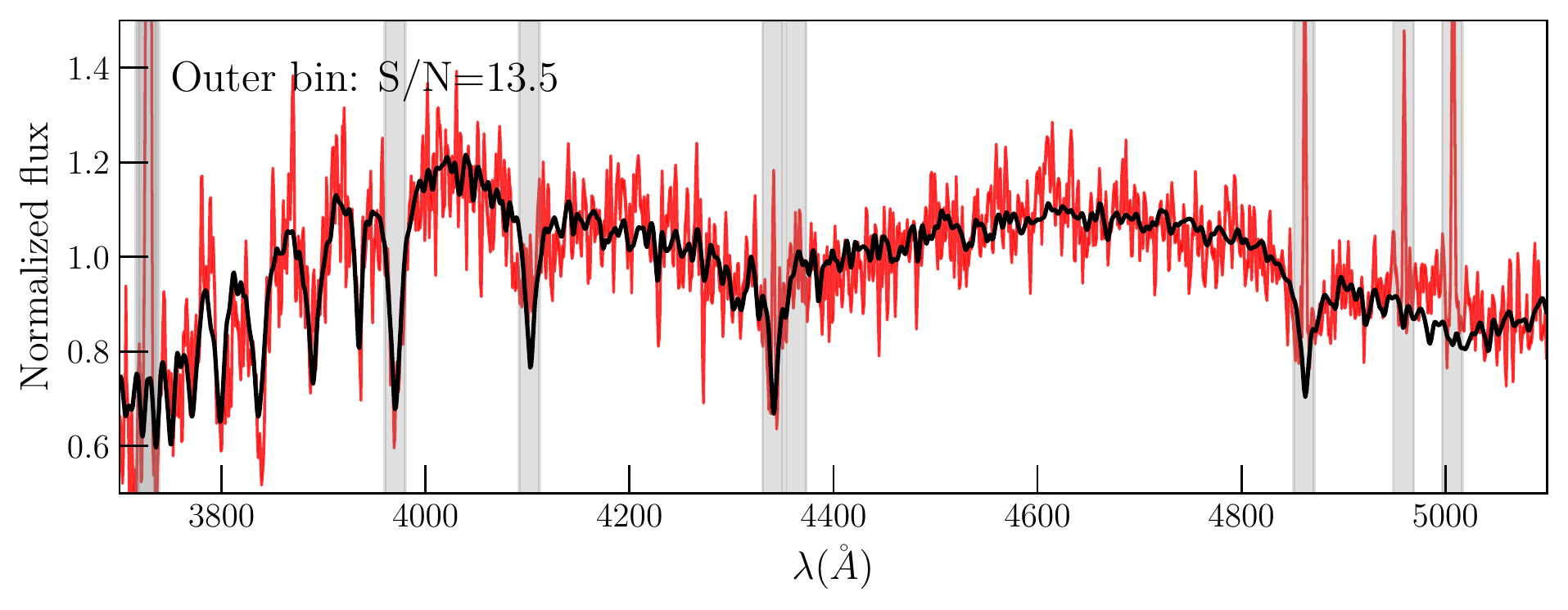}\\
     \plotone{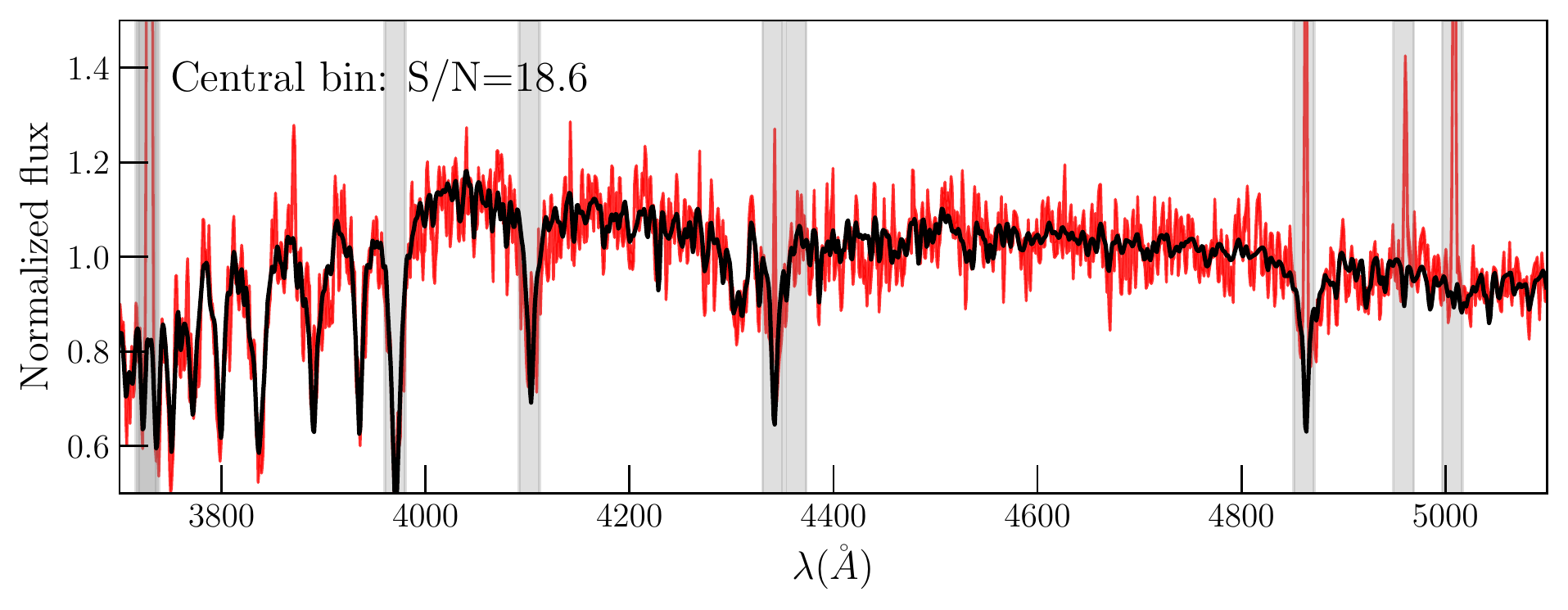}
     \plotone{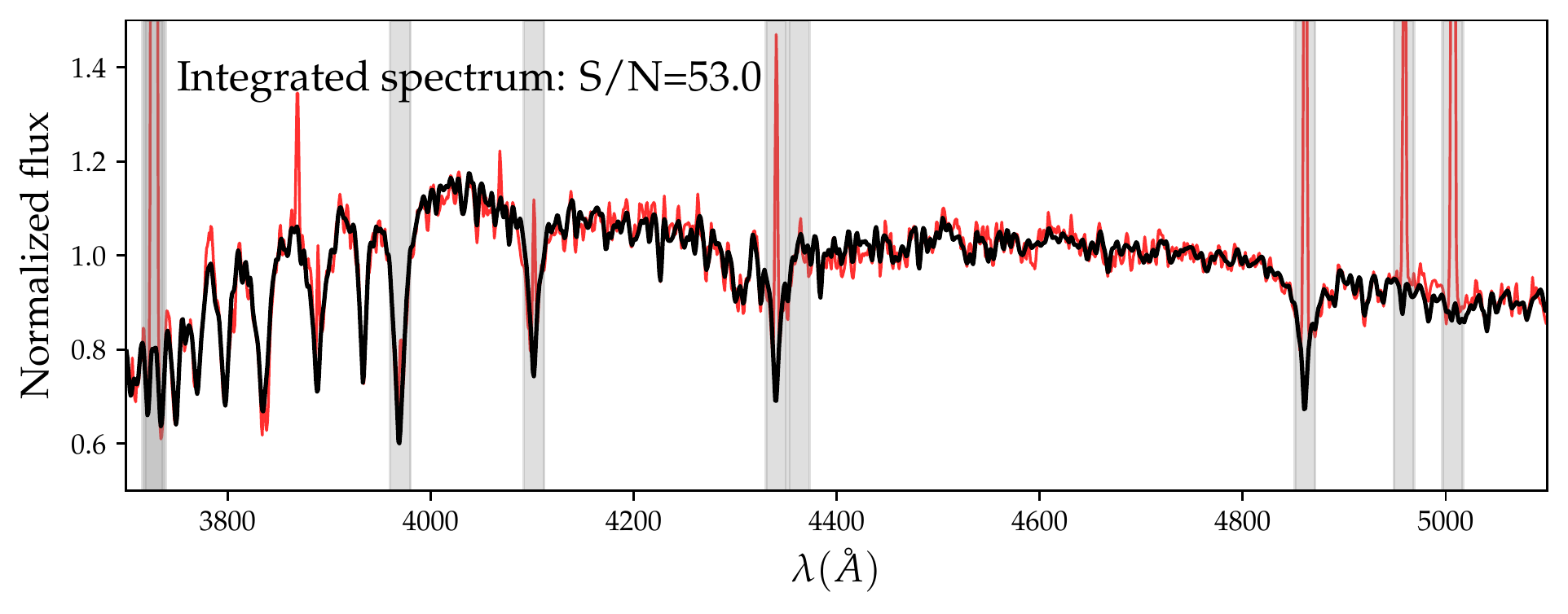}
    \caption{Example \texttt{pPXF} fits for binned and integrated spectra. Top: \texttt{pPXF} fit (black line) to observed spectrum (red line) for a bin in the outer region of field galaxy reines65, with relatively low continuum S/N (Equation~\ref{eq:sn-cont}). Gray shaded regions indicate gas emission lines, which are masked out of the spectrum before fitting. Middle: Same, but for a bin in the center of reines65, with higher S/N. Bottom: Same, but for the integrated spectrum, which has higher S/N than any of the binned spectra.}
    \label{fig:specfit}
\end{figure*}

The instrumental resolution of KCWI is slightly better, with a FWHM of $\sim2.4$\AA{} (as measured from arc lamps).
We correct this for redshift by dividing by $(1+z)$ \citep[see][]{Cappellari2017}, then use a Gaussian kernel to smooth the observed spectra to match the spectral resolution of the templates. 
Before using the \texttt{pPXF} algorithm, we also mask strong gas emission lines from the observed spectra and normalize both observed and template spectra by their median values.

For the $i^{\rm th}$ binned spectrum, \texttt{pPXF} recovers the line-of-sight stellar velocity $v_{i}$ and velocity dispersion $\sigma_{\star,i}$.
The top two panels of Figure~\ref{fig:specfit} illustrate \texttt{pPXF} fits to binned spectra with different S/N.
For comparison, the bottom panel shows the integrated (i.e., inverse variance-weighted average) spectrum.
\texttt{pPXF} is able to fit all spectra despite their varying S/N.
Any bins with $\mathrm{S/N}<1$ (primarily bins that are outside the galaxy) are discarded from the remainder of our analysis.
Additionally, velocity dispersion measurements $\sigma_{\star,i}$ are also limited by resolution uncertainties: for example, the instrumental dispersion of observed spectra, or mismatches between the resolutions of observed and template spectra \citep{Cappellari2017}.
As a result, in some bins \texttt{pPXF} may be able to recover accurate stellar velocities but not velocity dispersions.
We therefore also discard measurements of $\sigma_{\star,i}$ in bins where $\sigma_{\star,i}<1$~km/s, since these are likely to be driven by noise or resolution effects \citep[see, e.g., Sec. 7.4.3 of][]{Westfall2019}.
For two galaxies (void galaxies 821857 and 1904061), this criterion leaves $<10$ bins (located preferentially near the galaxy centers, which are typically the regions with the lowest $\sigma_{\star,i}$); given such a small number of bins, we do not report global velocity dispersions for these galaxies.
Finally, we also remove from our analysis two other galaxies (void galaxy 1074435 and field galaxy SDSS J0133+1342) which had poor wavelength solutions in most of their bins.

\begin{figure*}[t!]
     \epsscale{1.12}
     \plotone{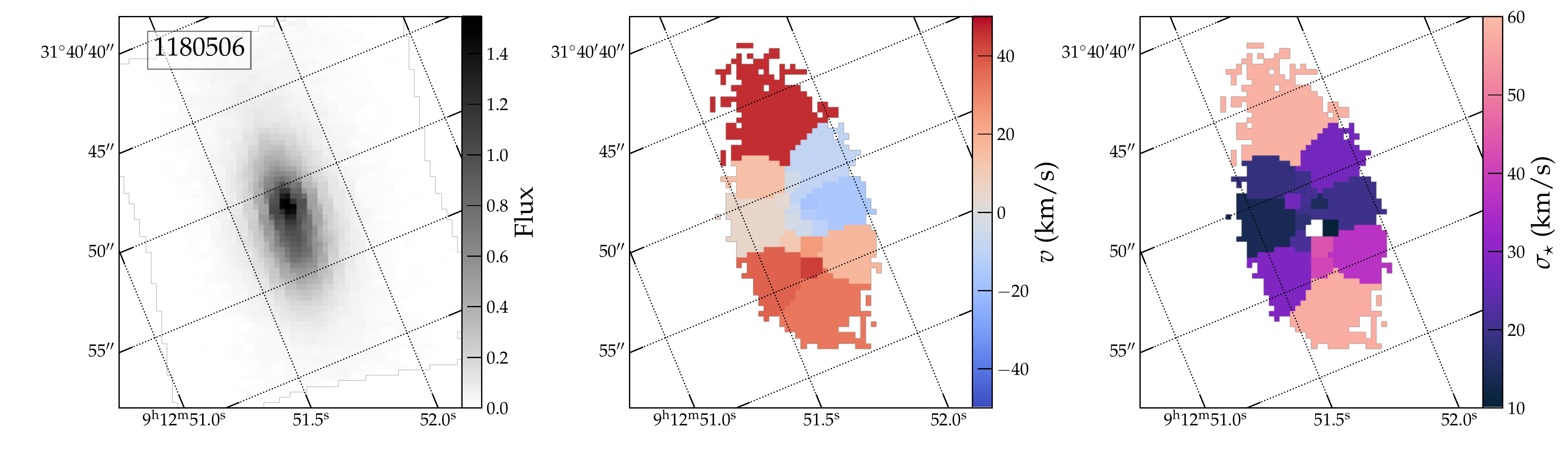}
     \plotone{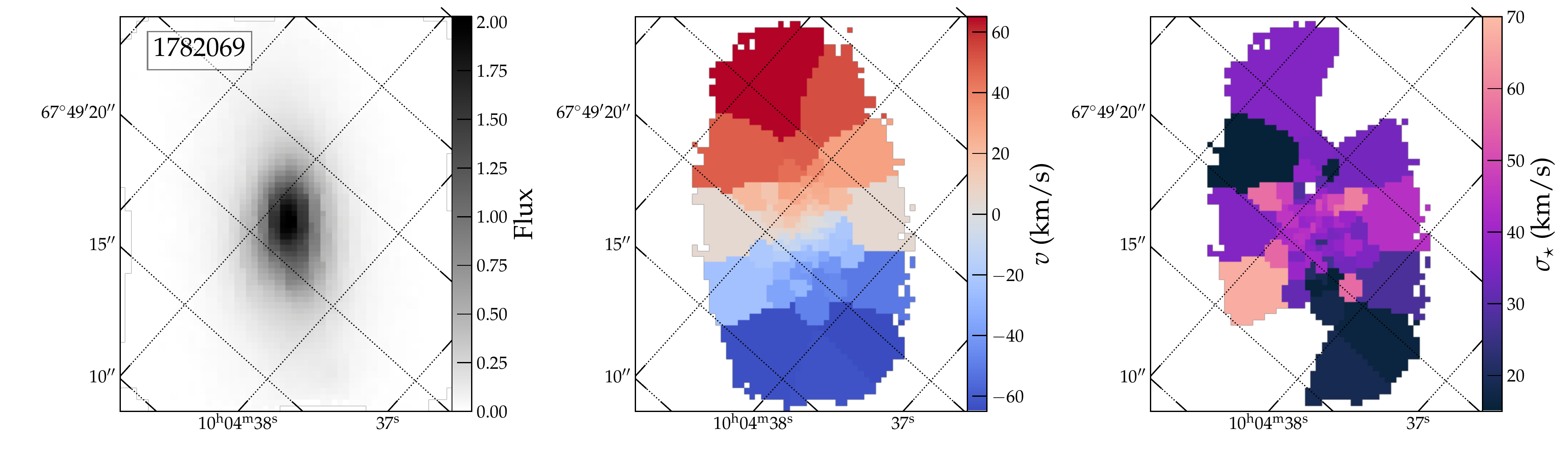}
     \plotone{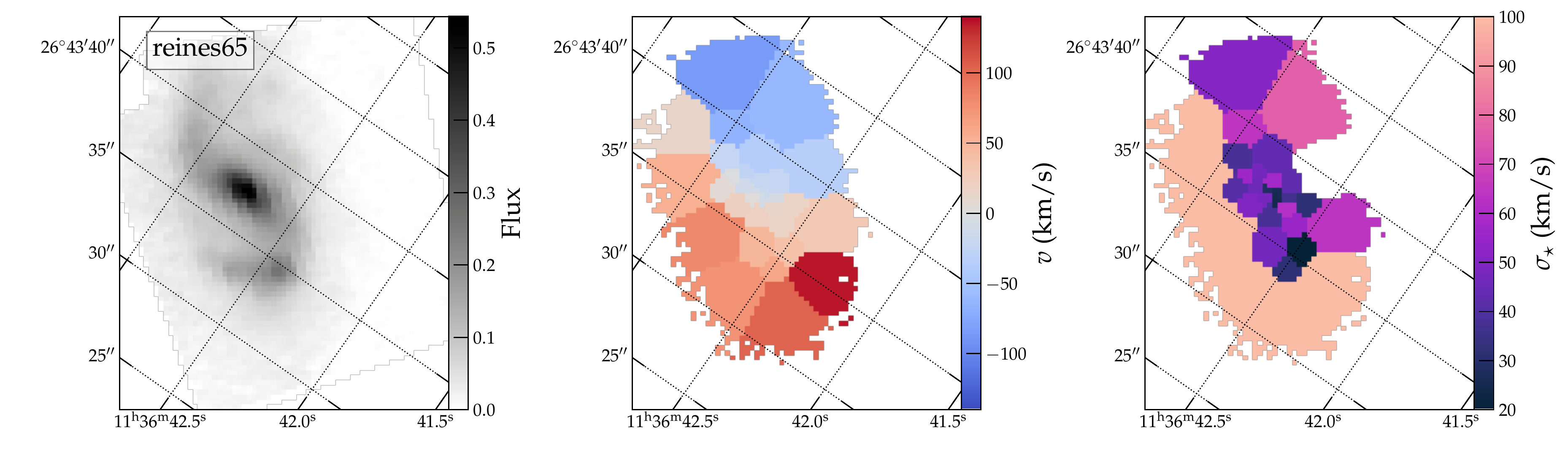}
     \plotone{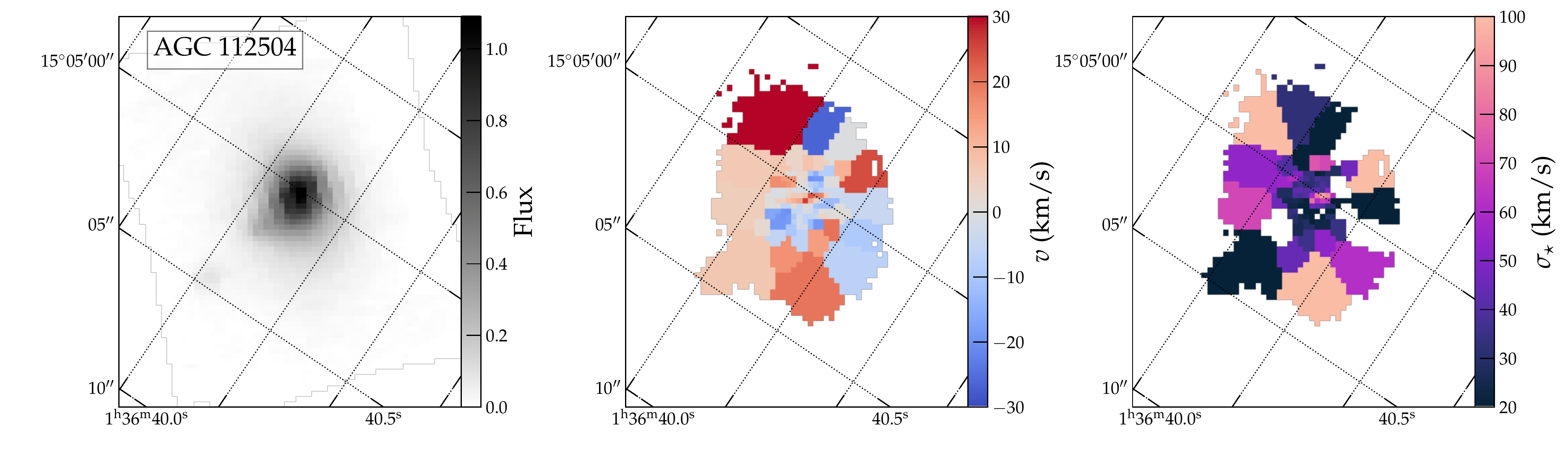}
    \caption{Example stellar kinematic maps of dwarf galaxies. Left: Integrated white-light images (in arbitrary flux units) computed by summing the flux in each IFU spaxel. Center: Maps of line-of-sight stellar velocity $v$ for dwarf galaxies. Right: Maps of line-of-sight stellar velocity dispersion $\sigma_{\star}$. As described in the text, bins with $\sigma_{\star,i}<1$~km/s are removed from our calculations, so we do not plot them here. Only some galaxies are presented in this figure; the complete figure set (11 images) is available in the online journal.}
    \label{fig:maps}
\end{figure*}

We compute the systemic line-of-sight velocity of the galaxy $v_{\mathrm{syst}}$, which we take to be the flux-weighted average of the velocities $v_{i}$ in each bin.
This systematic velocity is subtracted from each of the bin velocities $v_{i}$.
The resulting maps of stellar (line-of-sight) velocity and velocity dispersion are presented in Figure~\ref{fig:maps}.

These spatially-resolved kinematic maps are then used to compute \emph{global} kinematic estimates.
In particular, the global ratio $v_{\mathrm{rot}}/\sigma$ is often used to trace the structural properties of galaxies.
Here, $v_{\mathrm{rot}}$ is the maximum rotational speed within the galaxy, and $\sigma_{\star}$ is the intrinsic local velocity dispersion.
While models can be used to estimate these global parameters from spatially-resolved kinematic information, this is unfortunately not a straightforward task for the low-mass dwarf galaxies in our sample.
Most of the galaxies in our sample do not have clear velocity gradients and are likely not well-described by simple inclined disk models.
More sophisticated models, such as Jeans models, are frequently not well-suited for low-mass galaxies that are not necessarily in dynamical equilibrium \citep[][]{El-Badry2017}.

We therefore consider empirical kinematic statistics.
Several recent studies \citep[e.g.,][]{Fraser-McKelvie2022} have used the formalism developed by \citet{Cappellari2007} and \citet{Binney2005} for IFU data to estimate $v_{\mathrm{rot}}/\sigma_{\star}$:
\begin{equation}
    \left(\frac{v_{\mathrm{rot}}}{\sigma}\right)^{2} = \frac{\sum_{i}F_{i}v_{i}^2}{\sum_{i}F_{i}\sigma_{\star,i}^2},
    \label{eq:vsigma}
\end{equation}
where $F_{i}$ is the total flux in the $i^{\rm th}$ bin.
However, Equation~\ref{eq:vsigma} was initially developed to measure anisotropy in elliptical galaxies.
As \citet{Fraser-McKelvie2022} note, brighter central regions are likely to dominate this flux-weighted measurement, leading to lower values of $v_{\mathrm{rot}}/\sigma_{\star}$---an effect that may be especially problematic for non-elliptical galaxies.

Following a number of IFU studies \citep[e.g.,][]{Law2009}, we instead approximate $v_{\mathrm{rot}}$ for a galaxy using the following formula for shear velocity:
\begin{equation}
v_{\mathrm{rot}}=\frac{1}{2}(v_{\mathrm{max}}-v_{\mathrm{\mathrm{min}}}).
\end{equation}
To prevent outliers in $v_{i}$ from dominating $v_{\mathrm{rot}}$, we define $v_{\mathrm{max}}$ and $v_{\mathrm{min}}$ as the respective medians of the $95-100$th and $0-5$th percentiles of $v_{i}$ in the galaxy's velocity map \citep[][]{Herenz2016}.
Errors in $v_{\mathrm{max}}$ and $v_{\mathrm{min}}$ are computed as the ranges of the $95-100$th and $0-5$th percentile of $v_{i}$.
Here we take $v_{i}$ to be the line-of-sight velocity and ignore the geometric effect of inclination angle $i$; we discuss the impact of this effect in Section~\ref{sec:vsig_uncertainty}.
We then estimate the global stellar velocity dispersion $\sigma_{\star}$ as the flux-weighted average of the bin velocity dispersions $\sigma_{i}$ \citep{Binney2005}.
The final uncertainties in both $v_{\mathrm{rot}}$ and $\sigma_{\star}$ are then computed using standard error propagation.
Table~\ref{stellarkinematics} summarizes the measured global stellar kinematics (systemic velocities, peak rotational velocities, velocity dispersions, and our final $v_{\mathrm{rot}}/\sigma_{\star}$ estimates) for the dwarf galaxies in the sample.

\begin{deluxetable*}{lllllcll}
\tablecolumns{8} 
\tablecaption{Stellar kinematics and derived properties of void and field dwarf galaxies. \label{stellarkinematics}} 
\tablehead{ 
\colhead{Object} & \colhead{$v_{\mathrm{syst}}$} & \colhead{$v_{\mathrm{rot}}$} & \colhead{$\sigma_{\star}$} & \colhead{$v_{\mathrm{rot}}/\sigma_{\star}$} & $d_{L^{\star}}$ & $\log M{\star}$\tablenotemark{a} & \colhead{$i$}\\
\colhead{} & \colhead{(km/s)} & \colhead{(km/s)} & \colhead{(km/s)} & \colhead{} & \colhead{(kpc)} & \colhead{[$M_{\odot}$]} & \colhead{($^\circ$)}
}
\startdata
\multicolumn{8}{c}{Void dwarf galaxies}\\
\tableline
    1180506 & $35.36$ & $30.75\pm1.47$ & $23.10\pm2.96$ & $1.33\pm0.18$ & 3070.1 & $7.55^{+0.23}_{-0.23}$ & 65.5 \\
    281238 & $85.26$ & $39.88\pm13.24$ & $54.98\pm12.14$ & $0.73\pm0.29$ & 5284.1 & $6.88^{+0.38}_{-0.38}$ & 58.3 \\
    1904061\tablenotemark{b} & $67.02$ & $22.95\pm4.44$ & \ldots & \ldots & 3181.9 & $6.91^{+0.35}_{-0.36}$ & 20.3 \\
    821857\tablenotemark{b} & $89.80$ & $50.08\pm20.01$ & \ldots & \ldots & 5960.8 & $7.49^{+0.30}_{-0.30}$ & 59.4 \\
    1158932 & $45.07$ & $22.29\pm7.26$ & $39.62\pm12.30$ & $0.56\pm0.25$ & 3033.7 & $7.65 ^{+0.37}_{-0.36}$ & 69.0 \\
    866934 & $74.27$ & $38.37\pm13.96$ & $33.68\pm5.02$ & $1.14\pm0.45$ & 1955.5 & $8.36^{+0.20}_{-0.20}$ & 68.6 \\
    825059 & $83.29$ & $22.60\pm8.23$ & $32.08\pm3.15$ & $0.70\pm0.27$ & 2769.9 & $8.14^{+0.34}_{-0.35}$ & 63.1 \\
    2502521 & $64.56$ & $40.14\pm9.89$ & $38.60\pm10.40$ & $1.04\pm0.38$ & 1542.5 & $9.53^{+0.57}_{-0.58}$ & 62.7 \\
    1228631 & $55.78$ & $26.57\pm5.31$ & $46.95\pm4.11$ & $0.57\pm0.12$ & 3036.6 & $7.87^{+0.13}_{-0.14}$ & 57.3 \\
    1876887 & $94.82$ & $35.58\pm13.36$ & $23.58\pm6.03$ & $1.51\pm0.69$ & 3486.5 & $8.21^{+0.17}_{-0.17}$ & 57.7 \\
    1246626 & $51.66$ & $58.13\pm20.68$ & $70.15\pm19.78$ & $0.83\pm0.38$ & 5862.8 & $7.63^{+0.25}_{-0.26}$ & 58.5 \\
    1142116 & $123.18$ & $18.59\pm7.27$ & $56.24\pm1.32$ & $0.33\pm0.13$ & 8132.2 & $8.58^{+0.14}_{-0.13}$ & 47.2 \\
    955106 & $90.30$ & $16.75\pm2.42$ & $22.36\pm4.55$ & $0.75\pm0.19$ & 4416.1 & $8.64^{+0.09}_{-0.09}$ & 45.1 \\
    1063413 & $18.10$ & $52.65\pm18.34$ & $29.65\pm3.33$ & $1.78\pm0.65$ & 4795.5 & $8.66^{+0.21}_{-0.21}$ & 74.6 \\
    1074435\tablenotemark{c} & \ldots & \ldots & \ldots & \ldots & \ldots & \ldots & \ldots \\
    1785212 & $71.42$ & $38.66\pm17.93$ & $21.50\pm5.47$ & $1.80\pm0.95$ & 2238.2 & $8.78^{+0.17}_{-0.17}$ & 44.0 \\
    1280160 & $73.56$ & $14.65\pm3.88$ & $34.61\pm1.48$ & $0.42\pm0.11$ & 2449.1 & $8.59^{+0.17}_{-0.17}$ & 48.1 \\
    1782069 & $70.05$ & $58.95\pm16.46$ & $38.51\pm2.20$ & $1.53\pm0.44$ & 2440.1 & $8.82^{+0.11}_{-0.11}$ & 61.2 \\
    1126100 & $-11.55$ & $42.86\pm5.54$ & $34.92\pm1.20$ & $1.23\pm0.16$ & 847.4 & $9.32^{+0.09}_{-0.09}$ & 38.6 \\
    PiscesA & $98.44$ & $65.24\pm23.37$ & $139.19\pm16.46$ & $0.47\pm0.18$ & 4265.5 & \ldots & \ldots \\
    PiscesB & $120.78$ & $28.48\pm1.36$ & $144.35\pm48.10$ & $0.20\pm0.07$ & 1760.8 & \ldots & \ldots \\
\tableline
\multicolumn{8}{c}{Field dwarf galaxies}\\
\tableline
    SDSS J0133+1342\tablenotemark{c} & \ldots & \ldots & \ldots & \ldots & \ldots & \ldots & \ldots \\
	AGC 112504 & $115.55$ & $25.81\pm8.74$ & $39.22\pm5.22$ & $0.66\pm0.24$ & 7071.5 & $7.89^{+0.28}_{-0.29}$ & 36.7\\
	UM 240 & $-130.73$ & $20.26\pm3.31$ & $39.31\pm10.98$ & $0.52\pm0.17$ & 10227.2 & $7.48^{+0.33}_{-0.33}$ & 43.0\\
	SHOC 150 & $106.69$ & $22.38\pm1.14$ & $23.06\pm5.45$ & $0.97\pm0.23$ & 3948.7 & $7.20^{+0.20}_{-0.20}$ & 57.8\\
	LEDA 3524 & $101.15$ & $32.44\pm2.50$ & $35.08\pm8.28$ & $0.92\pm0.23$ & 1235.3 & $7.65^{+0.25}_{-0.26}$ & 66.1\\
	LEDA 101427 & $109.00$ & $16.74\pm6.16$ & $18.53\pm3.75$ & $0.90\pm0.38$ & 4287.5 & $8.62^{+0.19}_{-0.18}$ & 58.6\\
	IC 0225 & $123.46$ & $20.22\pm8.11$ & $31.87\pm2.66$ & $0.63\pm0.26$ & 1700.8 & $8.55^{+0.10}_{-0.10}$ & 30.2\\
	reines65 & $127.08$ & $93.65\pm24.93$ & $58.37\pm9.50$ & $1.60\pm0.50$ & 545.4 & $7.51^{+0.24}_{-0.25}$ & 57.0
\enddata
\tablenotetext{a}{Stellar masses are calculated as described in Section~\ref{sec:mass}.}
\tablenotetext{b}{As described in the text, we did not compute $\sigma_{\star}$ for these galaxies because they had $<10$ bins in which the velocity dispersion is large enough to be a reliable measurement.}
\tablenotetext{c}{As described in the text, the wavelength solutions for these galaxies are poor, so we do not consider their kinematic measurements in the rest of this paper.}
\end{deluxetable*}

\section{Discussion}
\label{sec:discussion}

The global value of $v_{\mathrm{rot}}/\sigma_{\star}$ is a probe of the stellar motions within a galaxy \citep[e.g.,][]{Illingworth1977,Bender1993,Binney2005,FerreMateu2021}.
Higher values of this ratio ($v_{\mathrm{rot}}/\sigma_{\star}\gg 1$) indicate dynamically cold, rotation-supported ``disky'' systems, while ``puffy'' dispersion-supported systems like dSphs typically have $v_{\mathrm{rot}}/\sigma_{\star}<1$.
The dependence of $v_{\mathrm{rot}}/\sigma_{\star}$ on external galaxy environment and on intrinsic galaxy properties is therefore a useful metric for understanding the processes that drive galaxy formation and dynamical evolution.

As shown in Table~\ref{stellarkinematics}, there is no significant difference in $v_{\mathrm{rot}}/\sigma_{\star}$ between the void and field subsamples: on average, the void dwarf galaxies have $\langle v_{\mathrm{rot}}/\sigma_{\star}\rangle=0.90^{+0.09}_{-0.09}$, while the ``control'' field galaxies have $\langle v_{\mathrm{rot}}/\sigma_{\star}\rangle=0.90^{+0.12}_{-0.12}$.
The similarity between the galaxies selected to be in voids and the randomly selected ``field'' galaxies suggests that the large-scale void environment does not significantly affect the stellar kinematics of dwarf galaxies.

In both subsamples, most of the dwarf galaxies (12/20 void dwarf galaxies and 6/7 field dwarf galaxies) are primarily dispersion-supported, with $v_{\mathrm{rot}}/\sigma_{\star}<1$.
All have $v_{\mathrm{rot}}/\sigma_{\star}<2$, implying that none of them is strongly rotating.
Both subsamples have slightly higher $v_{\mathrm{rot}}/\sigma_{\star}$ than Local Group dwarf galaxies on average \citep[$\langle v_{\mathrm{rot}}/\sigma_{\star}\rangle=0.50^{+0.09}_{-0.09}$;][]{Wheeler2017}.
In the following sections, we discuss potential physical implications for this discrepancy.

\subsection{Disk formation as a function of proximity to a massive galaxy}
\label{sec:dLstar}

We now consider the effect of local environment on dwarf galaxy kinematics. 
In particular, our $v_{\mathrm{rot}}/\sigma_{\star}$ measurements allow us to directly test the predictions of the tidal stirring model. 
In this model, which posits that tidal interactions remove angular momentum from dwarf galaxy disks during pericentric passages, $v_{\mathrm{rot}}/\sigma_{\star}$ is expected to increase with increasing distance from a massive galaxy \citep{Kazantzidis2011}.
\citet{Wheeler2017} did a systematic search of 40 Local Group dwarf galaxies and did not find evidence of a trend between $v_{\mathrm{rot}}/\sigma_{\star}$ and distance $d_{L^{\star}}$ to a massive host (i.e., the Milky Way or M31).
Since void dwarf galaxies are typically extremely isolated, they can be used to extend this analysis to higher $d_{L^{\star}}$.

We first use the SDSS DR17 catalog to locate the closest massive neighbor to each dwarf galaxy in our sample.
We identify all galaxies with apparent magnitudes $14 < g < 20$~mag and compute their stellar masses using mid-infrared photometry measured from the Wide-field Infrared Survey Explorer \citep[WISE;][]{Wright2010}.
We use the empirical relation from \citet{Cluver2014} for resolved low-redshift sources:
\begin{equation}
    \log_{10}M_{\star}/L_{\mathrm{W1}} = -2.54(W_{\mathrm{3.4~\mu m}} - W_{\mathrm{4.6~\mu m}}) - 0.17,
\label{eq:mass}
\end{equation}
where $L_{W1} (L_{\odot}) = 10^{-0.4(M_{W1}-3.24)}$.
Here, $M_{W1}$ is the absolute magnitude of the source in the WISE $3.4~\mu m$ band and $W_{\mathrm{3.4~\mu m}} - W_{\mathrm{4.6~\mu m}}$ is the rest-frame color.
These magnitudes and colors are obtained from the WISE ``forced photometry'' catalog produced for SDSS \citep{Lang2016}.
For each of the dwarf galaxies in our sample, we then compute three-dimensional distances to all massive ($M_{\star}>10^{10}~M_{\odot}$) neighbors, using SDSS coordinates and spectroscopic redshifts\footnote{To convert redshifts to distances, we assume a flat $\Lambda$CDM cosmology with Planck 2018 parameters \citep[$H_{0}=67.4~\mathrm{km}~\mathrm{s}^{-1}\mathrm{Mpc}^{-1}$, $\Omega_{m}=0.315$;][]{Planck2018}.} and identify the nearest massive neighbor.
The distances to the nearest massive galaxies, $d_{L^{\star}}$, are listed in Table~\ref{stellarkinematics}.

As expected, the void dwarf galaxies in our sample are extremely isolated, with $d_{L^{\star}}\gtrsim1$~Mpc. 
Interestingly, most of the randomly selected field galaxies also appear to have similarly high values of $d_{L^{\star}}$.
This is not entirely surprising: as discussed in Section~\ref{sec:sample}, the field dwarf galaxies could not be directly compared against the void catalogs used to select the void galaxies, and some of them may in fact be located in large-scale void environments.
Furthermore, we emphasize that $d_{L^{\star}}$ is a measure of \emph{local} isolation that may not necessarily correspond with large-scale (i.e., void-scale) environment; while void galaxies are expected to have high $d_{L^{\star}}$, galaxies with high $d_{L^{\star}}$ may not necessarily be in voids.

\begin{figure*}[t!]
     \epsscale{1}
     \plotone{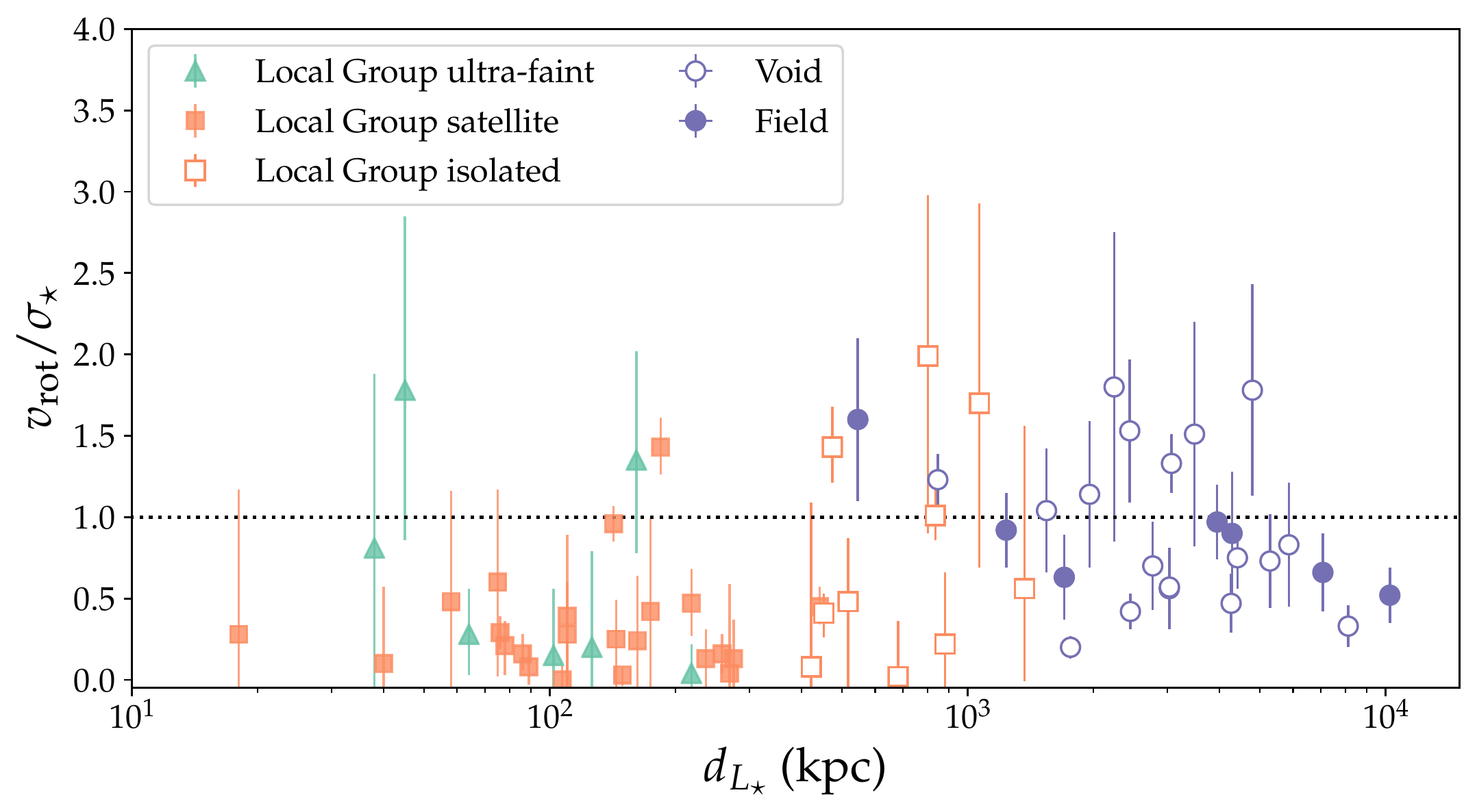}
     \plotone{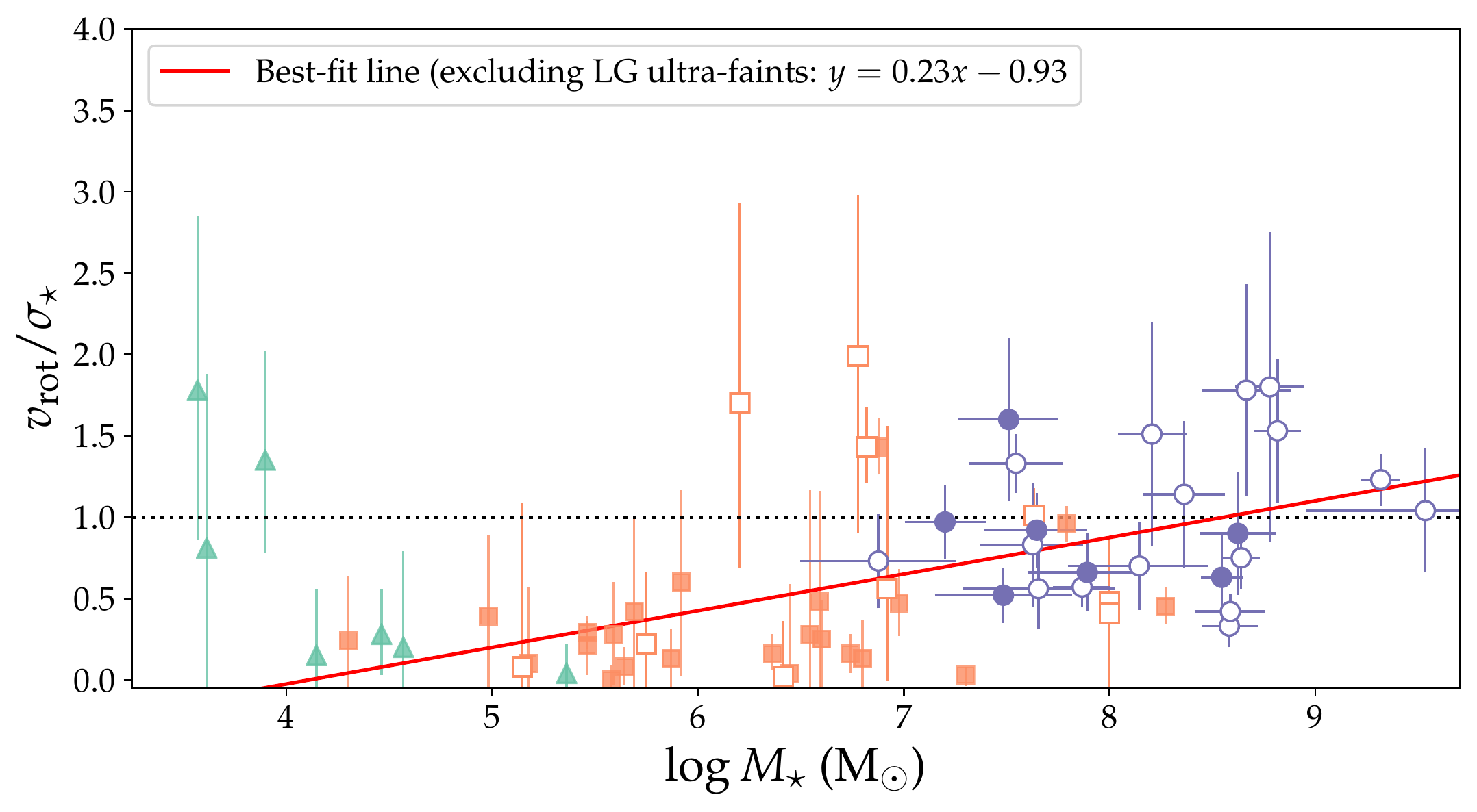}
    \caption{Stellar rotation support ($v_{\mathrm{rot}}/\sigma_{\star}$) as a function of distance from closest massive galaxy $d_{L^{\star}}$ (top) and stellar mass (bottom). Measurements for Local Group galaxies---ultra-faint dwarf galaxies (green triangles), MW and M31 satellites (filled orange squares), and isolated dwarf galaxies (open orange squares)---are taken from the compilation of \citet{Wheeler2017}. Measurements from this work are denoted as purple circles: filled circles denote ``field'' galaxies from the control sample, while open circles denote void galaxies.}
    \label{fig:vsigma}
\end{figure*}

The top panel of Figure~\ref{fig:vsigma} plots $v_{\mathrm{rot}}/\sigma_{\star}$ as a function of $d_{L^{\star}}$.
The void and field dwarf galaxies in our sample are plotted as purple empty and filled circles, respectively, while \citet{Wheeler2017}'s measurements of Local Group satellite (dSph) and isolated (dIrr) galaxies are respectively plotted as orange filled and empty squares.
A small number of ultra-faint dwarf galaxies, also measured by \citet{Wheeler2017}, are shown as filled green triangles for comparison.

The void and field dwarf galaxies in our sample have higher $d_{L^{\star}}$ than most of the Local Group galaxies, and they are located well outside the virial radii of any massive halos (for the Milky Way and M31, $R_{\mathrm{vir}}\sim300$~kpc).
Some studies have shown that dark matter halos can produce environmental effects at even greater radii---for example, so-called ``backsplash'' dwarf galaxies may have been within the virial radius of a host galaxy at earlier times in their orbits, despite currently being located outside $R_{\mathrm{vir}}$ \citep[e.g.,][and references therein]{Buck2019}.
Almost all the dwarf galaxies in our sample are located far enough from massive galaxies that they are unlikely to be backsplash galaxies \citep[see, e.g.,][who predict a ``splashback'' radius of $\sim2.5R_{\mathrm{vir}}$, or $\sim750$~kpc for Milky Way-like galaxies]{More2015}.

Additionally, in tidal stirring models, dIrrs require $>10$~Gyr of repeated pericentric passages within a massive host's gravitational potential to complete the full conversion to dSphs \citep[e.g., Table 3 of][]{Kazantzidis2017}.
In order to have had such prolonged tidal interactions in the past, many of the dwarf galaxies in our sample would have had to move extremely far from their massive hosts in a short time.
For a galaxy with $d_{L^{\star}}\sim3000$~kpc to have left its host within the last $3$~Gyr, it would have needed to move away from its host with an average speed of $\sim1000$~km~s$^{-1}$; for comparison, the LMC and SMC are estimated to be infalling on the Milky Way with velocities on the order of $\sim300$~km~s$^{-1}$ \citep{Kallivayalil2013}.
We therefore expect our sample to consist of dwarf galaxies that have not been affected by environmental processes like tidal stirring.

Assuming our sample does represent truly isolated dwarf galaxies, the relatively low values of $v_{\mathrm{rot}}/\sigma_{\star}<2$ that we measure are consistent with the results of some analytic calculations \citep[][]{Kaufmann2007} and hydrodynamic simulations \citep[][]{Wheeler2017,Frings2017}, which find that isolated dwarf galaxies naturally form as dispersion-supported systems.
However, our measurements are at odds with previous studies of tidal stirring, which typically assume that isolated dwarf galaxies are rotationally supported disks with $v_{\mathrm{rot}}/\sigma_{\star}\approx2$ \citet{Kazantzidis2017}.
Furthermore, as shown in the top panel of Figure~\ref{fig:vsigma}, there is no strong trend between $v_{\mathrm{rot}}/\sigma_{\star}$ and $d_{L^{\star}}$, as would be expected from tidal stirring models.
We conclude that tidal interactions are not needed to produce dispersion-supported dwarf galaxies.

\subsection{Disk formation as a function of stellar mass}
\label{sec:mass}

If, as our measurements suggest, low-mass galaxies are dispersion-supported stellar systems even in extreme isolation, perhaps they in fact form as ``puffy'' systems rather than disks.
Indeed, isolated dwarf galaxies in zoom-in hydrodynamic simulations appear to have low stellar $v_{\mathrm{rot}}/\sigma_{\star}$ without being subjected to external perturbations \citep[][]{Wheeler2017}.
Yet dynamically cold disks are common among more massive galaxies, perhaps implying that there is a critical stellar mass required to form a disk.
Some studies have estimated such a critical mass using morphological observations; for example, \citet{SanchezJanssen2010} find that galaxies below $M_{\star}=2\times10^{9}~M_{\odot}$ appear substantially thicker in terms of observed axial ratios $b/a$.
We now aim to expand on this work by using stellar kinematics, rather than morphological criteria, to search for a transition between dispersion-supported dwarfs and rotation-dominated massive galaxies.

We compute stellar masses for our galaxies using WISE photometry, following Equation~\ref{eq:mass}.
The mid-infrared magnitudes are obtained directly from the WISE All-Sky Data Release Source Catalog \citep[][]{Cutri2012}; we find that using the \citet{Lang2016} ``forced photometry'' SDSS catalog, as we did for the massive neighbor galaxies, does not significantly affect our results.
Table~\ref{stellarkinematics} lists the stellar masses for our sample. 
We consider the effect of different stellar mass calibrations in Section~\ref{sec:uncertainty_secondary}.

The bottom panel of Figure~\ref{fig:vsigma} illustrates the relation between $v_{\mathrm{rot}}/\sigma_{\star}$ and stellar mass.
\citet{Wheeler2017} found no clear trend between $v_{\mathrm{rot}}/\sigma_{\star}$ and $M_{\star}$ for Local Group dwarf galaxies across the mass range $M_{\star}=10^{3.5}-10^{8}~M_{\odot}$.
Our sample (purple circles) extends this relation to higher masses.
We see a slight upward trend in $v_{\mathrm{rot}}/\sigma_{\star}$, particularly above $M_{\star}\sim10^{7}~M_{\odot}$.
To quantify this trend, we use simple linear regression, which yields the best-fit line 
\begin{equation}
v_{\mathrm{rot}}/\sigma_{\star}=0.23^{+0.05}_{-0.05}(\log M_{\star}/M_{\odot})-0.93^{+0.37}_{-0.38}.
\label{eq:vsigma_mass}
\end{equation}
We exclude the ultra-faint dwarf galaxies from this fit, since their relatively high $v_{\mathrm{rot}}$ are likely the result of tidal distortion rather than coherent rotation \citep[][]{Simon2007}.

Despite this upward trend, it is evident that the low-mass galaxies plotted in Figure~\ref{fig:vsigma} are kinematically distinct from the thin disks seen in high-mass galaxies (for comparison, the Milky Way has $v_{\mathrm{rot}}/\sigma_{\star}\sim20$).
Measurements of stellar $v_{\mathrm{rot}}/\sigma_{\star}$ at higher stellar masses ($M_{\star}\gtrsim10^{9}~M_{\odot}$) are needed to identify whether the transition between dispersion-supported dwarfs and rotation-dominated massive galaxies is a gradual trend or a sharp discontinuity.

\subsection{Limitations of this study}
We now consider potential limitations of this study, including sources of errors that may affect our results.

\subsubsection{Selection effects}
Our sample is not mass complete, and our selection may bias our results.
As described in Section~\ref{sec:sample}, we preferentially select dwarf galaxies with high [\oiii]$\lambda 4363$ fluxes.
This would bias our sample towards galaxies with lower gas-phase metallicities at a given stellar mass.
At a given stellar mass, gas-phase metallicity may be inversely correlated with specific star formation rate \citep[the ``fundamental metallicity relation''; see, e.g., Section 5.2 of][and references therein]{Maiolino2019}, which may in turn be inversely correlated with velocity dispersion \citep[e.g.,][although note that these studies have focused on \emph{gas} kinematics in galaxies more massive ($>10^{9}~M_{\odot}$) than our sample]{Wake2012,vanderWel2016}.

Assuming these correlations hold for our sample, the dwarf galaxies in our sample may therefore be biased toward lower velocity dispersions---and higher $v_{\mathrm{rot}}/\sigma_{\star}$---at a given stellar mass.
In other words, a more representative sample of dwarf galaxies at similar masses might have \emph{lower} average $v_{\mathrm{rot}}/\sigma_{\star}$ values than our sample.
This further strengthens our main finding that isolated dwarf galaxies are dispersion-supported systems with low $v_{\mathrm{rot}}/\sigma_{\star}$.
This would also decrease the discrepancy between Local Group dwarf galaxies and isolated dwarf galaxies and weaken the trend between $v_{\mathrm{rot}}/\sigma_{\star}$ and $M_{\star}$, although the potential magnitude of this effect is difficult to estimate.

\subsubsection{Uncertainties in secondary parameters}
\label{sec:uncertainty_secondary}

There are a number of systematic effects that may impact our measurements of the galaxy properties $d_{L^{\star}}$ and $M_{\star}$.
First, our $d_{L^{\star}}$ estimates rely on SDSS redshifts.
Not only is there intrinsic scatter in the Hubble relation, but SDSS spectroscopic redshifts also have an uncertainty of $\Delta(cz)\sim30$~km/s \citep{Abazajian2005}.
Combined, these can correspond to distance uncertainties of up to thousands of kpc.
We test the effect of these uncertainties by randomly perturbing our $d_{L^{\star}}$ measurements (assuming a normal distribution with a standard deviation $1000$~kpc).
We find that our main qualitative result---that there is no strong trend between $v_{\mathrm{rot}}/\sigma_{\star}$ and $d_{L^{\star}}$---does not change.

\begin{figure}[t!]
     \epsscale{1.2}
     \plotone{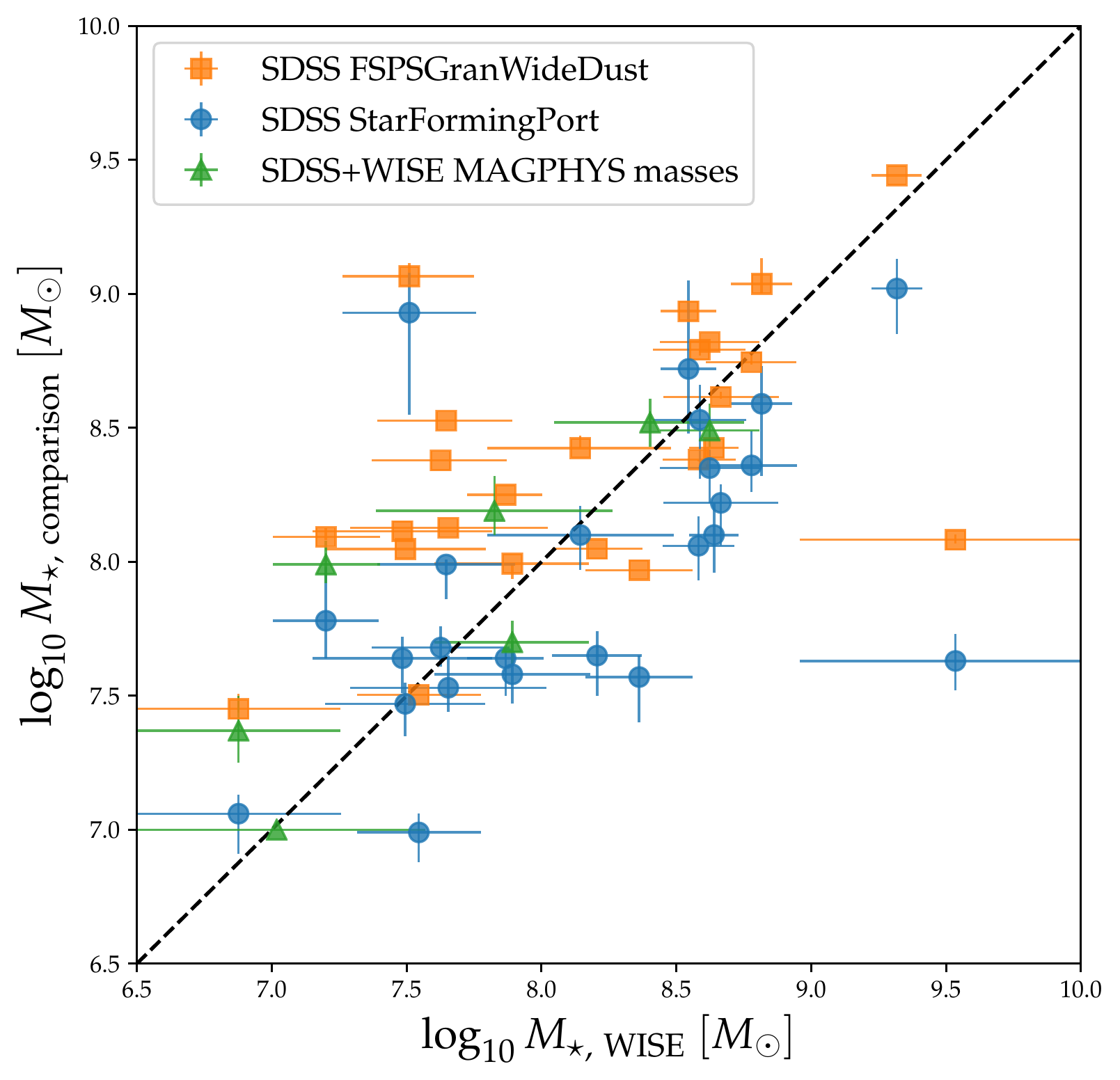}
    \caption{Comparison between stellar masses measured from WISE ($x$-axis) and stellar masses measured from other calibrations ($y$-axis). These include the Portsmouth calibration \citep[blue circles;][]{Maraston2005}, the Granada calibration \citep[orange squares;][]{Conroy2009}, and the SDSS+WISE MAGPHYS catalog \citep[green triangles;][]{Chang2015}.}
    \label{fig:masscomparison}
\end{figure}

The stellar masses are also uncertain. 
In Section~\ref{sec:mass} we chose to compute masses using available mid-infrared photometry from WISE, which is ideal for tracing old stellar populations that may be outshone by younger stars in the optical \citep[][]{Jarrett2013}.
We now compare these WISE masses with optical-based stellar mass calibrations in Figure~\ref{fig:masscomparison}.
We first consider two stellar mass estimates that are available as SDSS value-added catalogs.
The star-forming Portsmouth method (``StarFormingPort,'' blue circles) fits SDSS photometry with a stellar evolution model comprised of a metallicity and one of three star-formation histories \citep[][]{Maraston2005}. 
The Granada method (``FSPSGranWideDust,'' orange squares) uses the Flexible Stellar Population Synthesis models of \citet{Conroy2009} to fit SDSS photometry; we use the WideDust version, which allows for extended star formation histories and accounts for dust extinction. 
Finally, we obtain stellar masses from \citet{Chang2015}, who use the MAGPHYS algorithm to fit combined SDSS+WISE photometry.\footnote{Although we searched other catalogs, including the GALEX-SDSS-WISE Legacy Catalog \citep{Salim2016}, we did not find any overlap between these catalogs and our sample.}

We find significant scatter between the different stellar mass estimates, reflecting the many assumptions inherent in modeling galaxy spectral energy distributions \citep[][]{Conroy2013}.
The Portsmouth method yields stellar masses that are systematically lower than the WISE masses (rms deviation of $\sim0.6$~dex), while the Granada method yields stellar masses that are systematically higher than the WISE masses (with similar rms deviation of $\sim0.6$~dex).
Only a small number of galaxies in our sample have $M_{\star}$ measurements in the MAGPHYS catalog, so it is difficult to make quantitative comparisons.
However, in Figure~\ref{fig:masscomparison} the MAGPHYS masses are qualitatively more consistent with the WISE masses than the other SDSS calibrations are---perhaps not surprising, since the MAGPHYS catalog uses WISE data.
Regardless of overall agreement (or lack thereof) between these stellar mass calibrations, we find that using different calibrations does not impact our results; in particular, the best-fit trend in Equation~\ref{eq:vsigma_mass} does not change within parameter uncertainties.

\subsubsection{Uncertainties in $v_{\mathrm{rot}}/\sigma_{\star}$}
\label{sec:vsig_uncertainty}

While our qualitative conclusions are not significantly affected by systematic uncertainties on the $x$-axes of Figure~\ref{fig:vsigma}, additional systematic effects may result from our choice of kinematic statistic on the $y$-axis.
As we have defined them, $v_{\mathrm{rot}}$ and $\sigma_{\star}$ are not perfect tracers of stellar rotation and dispersion.

For example, if a galaxy's angular momentum vector is inclined relative to the line-of-sight by angle $i$, then the line-of-sight $v_{\mathrm{rot}}$ we measure is a lower limit to the intrinsic velocity: $v_{\mathrm{rot}} = v_{\mathrm{rot}}^{\mathrm{intrinsic}}\sin i$.
We can correct for this effect by estimating inclination angles $i$ from photometric axis ratios: $i=\cos^{-1}(b/a)$, where $a$ and $b$ are the semi-major and semi-minor axes measured by SDSS $r$-band exponential model fits.
We report $i$ in Table~\ref{stellarkinematics} for completeness.

Applying these first-order inclination corrections to our sample increases $v_{\mathrm{rot}}/\sigma_{\star}$ for all galaxies in our sample.
Three galaxies become more strongly disk-like (i.e., consistent with $v_{\mathrm{rot}}^{\mathrm{intrinsic}}/\sigma_{\star}\gtrsim2$): the $v_{\mathrm{rot}}/\sigma_{\star}$ for void galaxy 1785212 changes from 1.80 to 2.59, for void galaxy 1126100 from 1.23 to 1.97, and for field galaxy reines65 from 1.60 to 1.91.
These corrections also slightly increase the overall discrepancy between $v_{\mathrm{rot}}^{\mathrm{intrinsic}}/\sigma_{\star}$ of Local Group dwarf galaxies and our isolated dwarf galaxies.
The correlation between $v_{\mathrm{rot}}/\sigma_{\star}$ and $M_{\star}$, for example, becomes stronger:
\begin{equation}
v_{\mathrm{rot}}/\sigma_{\star}=0.31^{+0.05}_{-0.05}(\log M_{\star}/M_{\odot})-1.43^{+0.38}_{-0.38}.
\end{equation}

On the other hand, many of our conclusions do not change even after correcting for inclination.
The void and field subsamples still have similar $v_{\mathrm{rot}}^{\mathrm{intrinsic}}/\sigma_{\star}$ values.
Several of the galaxies in our sample still have relatively low $v_{\mathrm{rot}}^{\mathrm{intrinsic}}/\sigma_{\star}<1$, and most remain below $v_{\mathrm{rot}}^{\mathrm{intrinsic}}/\sigma_{\star}\sim2$, which is the value frequently used as an initial condition in tidal stirring simulations.
We continue to conclude that environmental effects are not required to produce ``puffy'' stellar systems.
Finally, the corrections described above may not be appropriate for galaxies that are not inclined disks---and as the line-of-sight velocity maps make clear, most dwarf galaxies in our sample have irregular morphologies and do not display signs of any coherent velocity structure that would be expected in a disk.

Another intrinsic limitation of our kinematic measurements is that although they are derived from spatially resolved IFU data, they are not actually measurements of resolved stars.
Comparisons with measurements of galaxies in the Local Group---which \emph{have} been derived from individual stellar velocities---are therefore not direct ``apples-to-apples'' comparisons.
Some studies have begun bridging this gap in recent years by comparing resolved stellar measurements with IFU or integrated-light measurements for individual dwarf galaxies.
For example, \citet{Ruiz-Lara2018} obtained both resolved stellar photometry and an integrated-light long-slit spectrum for Local Group dwarf galaxy Leo A and found that both measurement techniques yielded consistent star formation histories.
Similarly, \citet{Zhuang2021} found that stellar metallicity estimates from integrated light were consistent with measurements from resolved stellar spectroscopy for M31 satellite NGC 147.
To our knowledge such a comparison for stellar kinematics measurements has not yet been done in one galaxy---let alone a systematic analysis of a population of galaxies.
In light of these observational obstacles, simulations may be the most promising tool to address this question, since they could be used to produce mock observations in order to quantify systematic differences between resolved and IFU observations.

\section{Conclusions}
\label{sec:conclusion}

Using the Keck Cosmic Web Imager, we have obtained IFU spectroscopy for a number of dwarf galaxies ($M_{\star}=10^{7}-10^{9}~M_{\odot}$) located inside and outside of cosmic voids.
In this work, we investigated the stellar dynamics of these galaxies by measuring their spatially resolved stellar velocities and stellar velocity dispersions.
From these spatially resolved measurements, we estimated global values of $v_{\mathrm{rot}}/\sigma_{\star}$, a parameter that probes the ratio of stellar rotation to dispersion.
Our findings are as follows.

\begin{enumerate}
    \item We find no significant difference between $v_{\mathrm{rot}}/\sigma_{\star}$ for the dwarf galaxies in cosmic voids and the ``field'' dwarf galaxies, suggesting that isolated dwarf galaxies have similar stellar dynamics regardless of large-scale environment.
    \item Environmental processes in general do not appear to be primary drivers of the dynamical formation of field dwarf galaxies. The majority of dwarf galaxies in our sample are, like Local Group dwarf galaxies, predominantly dispersion-supported ($v_{\mathrm{rot}}/\sigma_{\star}\lesssim1$), and we find no correlation between $v_{\mathrm{rot}}/\sigma_{\star}$ and distance to the closest massive galaxy $d_{L^{\star}}$. This confirms the results of \citet{Wheeler2017}, who find a similar lack of trend among Local Group dwarf galaxies. These results are further evidence that dSphs/dEs form as ``puffy'' dispersion-supported systems, rather than as rotation-supported disks that are converted by tidal interactions into dispersion-supported spheroidal systems (the ``tidal stirring'' hypothesis).
    \item We find evidence that $v_{\mathrm{rot}}/\sigma_{\star}$ begins to increase with galaxy stellar mass at $M_{\star}\gtrsim10^{7}~M_{\odot}$. However, measurements at higher $M_{\star}$ are needed to identify the transition to strongly rotation-dominated ($v_{\mathrm{rot}}/\sigma_{\star}\gg2$) disks in higher-mass galaxies.
    \item We consider a number of limitations of our study, including uncertainties on $d_{L^{\star}}$, $M_{\star}$, and $v_{\mathrm{rot}}/\sigma_{\star}$. We find that while known uncertainties in $d_{L^{\star}}$ and $M_{\star}$ are unlikely to impact our results, $v_{\mathrm{rot}}/\sigma_{\star}$ may be subject to a number of systematic effects that could modify some of our conclusions (for example, inclination effects may strengthen a correlation between $v_{\mathrm{rot}}/\sigma_{\star}$ and $M_{\star}$). Yet even accounting for these effects, many of the galaxies in our sample still have $v_{\mathrm{rot}}/\sigma_{\star}<2$. Our most conservative conclusion is therefore that environmental effects are not strictly required to make dispersion-supported, low-mass galaxies.
\end{enumerate}

Further investigation is needed to understand other potential systematic effects, particularly the effect of spatial resolution. 
In other words, are stellar kinematic measurements obtained from resolved stellar populations (as for Local Group dwarf galaxies) systematically different from stellar kinematic measurements obtained from IFU data (as this paper does)?
This question also represents a broader issue for the astrophysical community: the Local Group has long served as a benchmark for theories of low-mass galaxy formation and evolution, yet the methods we use to observe Local Group galaxies are very different from those used to study more distant galaxies.
As we begin to explore statistical populations of low-mass galaxies beyond the Local Group, it is critical to understand how differences in observational techniques can impact our measurements. 
Future simulations and observations can provide a path forward.

In the meantime, integral field spectroscopy remains one of our most promising tools for obtaining spatially resolved observations of low-mass galaxies beyond the Local Group.
As this study demonstrates, dwarf galaxies located in cosmic voids are a particularly interesting sample for targeting with IFU surveys, since they can probe in-situ physical processes by minimizing the effect of environment.
Measuring stellar kinematics is just one of the many applications of the KCWI dataset presented in this study.
In future work, we will investigate other properties that are observable from these rich data; in particular, ionized gas kinematics will provide an useful comparison to the stellar kinematics presented here.

\acknowledgments
{This material is based upon work supported by the National Science Foundation under grant No. AST-2233781. 
MAdlR acknowledges the financial support of the NSF Graduate Research Fellowship and the Stanford Science Fellowship. 
ZZ acknowledges the financial support of the NASA FINESST Program (No.\ 80NSSC22K1755).
CCS has been supported in part by NSF AST-2009278.

This research has made use of NASA’s Astrophysics Data System Bibliographic Services.

There are many communities without whom this work would not have been possible. 
We acknowledge that this work is rooted in Western scientific practices and is the material product of a long and complex history of settler-colonialism. MAdlR and ENK wish to recognize their status as settlers on the ancestral lands of the Muwekma Ohlone Tribe and the Potawatomi people, and to recognize that the astronomical observations described in this paper were only possible because of the dispossession of Maunakea from K$\bar{\mathrm{a}}$naka Maoli. We hope to work toward a scientific practice guided by pono and a future in which we all honor the land.}

\vspace{5mm}
\facilities{Keck:II (KCWI)}

\software{
Matplotlib \citep{matplotlib}, 
Astropy \citep{astropy},
Scipy \citep{scipy},
CWITools \citep{OSullivan2020}}

\bibliographystyle{aasjournal}
\bibliography{void}

\end{document}